\documentclass[twocolumn]{aastex7}

\def\HI{H\,{\textsc{\romannumeral 1}}}

\usepackage{longtable} 
\begin{document}

\title[large]{HI-detected Dwarf Galaxies in the FASHI Survey: Insights from Single- and Double-Peaked Emission-Line Samples}

\author[orcid=0000-0003-0202-0534]{Cheng Cheng}
\affiliation{Chinese Academy of Sciences South America Center for Astronomy, National Astronomical Observatories, CAS, Beijing 100101, China} 
\affiliation{Key Laboratory of Optical Astronomy, NAOC, 20A Datun Road, Chaoyang District, Beijing 100101, China}
\email[show]{chengcheng@nao.cas.cn}

\author[0000-0001-6511-8745]{Jia-Sheng Huang}
\affiliation{Chinese Academy of Sciences South America Center for Astronomy, National Astronomical Observatories, CAS, Beijing 100101, China}
\affiliation{Harvard-Smithsonian Center for Astrophysics, 60 Garden Street, Cambridge, MA 02138, USA}
\email{jhuang@nao.cas.cn}

\author[0000-0003-4546-8216]{Wei Du}
\affiliation{National Astronomical Observatories, Chinese Academy of Sciences, Beijing 100101, People's Republic of China}
\affiliation{Key Laboratory of Optical Astronomy, NAOC, 20A Datun Road, Chaoyang District, Beijing 100101, China}
\email{wdu@nao.cas.cn}

\author[0000-0003-1632-2541]{Hong-Xin Zhang}
\affiliation{Department of Astronomy, University of Science and Technology of China, Hefei, Anhui 230026, People's Republic of China}
\affiliation{School of Astronomy and Space Sciences, University of Science and Technology of China, Hefei 230026, Anhui, People's Republic of China}
\email{hzhang18@ustc.edu.cn}

\author[0000-0002-4428-3183]{Chuan-Peng Zhang}
\affiliation{National Astronomical Observatories, Chinese Academy of Sciences, Beijing 100101, People's Republic of China}
\affiliation{Guizhou Radio Astronomical Observatory, Guizhou University, Guiyang 550000, People's Republic of China}
\email{cpzhang@nao.cas.cn}

\author[0000-0001-6083-956X]{Ming Zhu}
\affiliation{National Astronomical Observatories, Chinese Academy of Sciences, Beijing 100101, People's Republic of China}
\affiliation{Guizhou Radio Astronomical Observatory, Guizhou University, Guiyang 550000, People's Republic of China}
\email{mz@nao.cas.cn}

\author[0000-0002-6642-7483]{Gustavo Orellana}
\affiliation{Fundaci\'on Chilena de Astronom\'ia, c\'odigo postal 7500011, Santiago, Chile}
\email{gustavo.orellana.gonzalez@gmail.com}

\begin{abstract}
We present a sample of low \HI\ mass ($M_{\rm HI} < 10^8 M_\odot$) dwarf galaxies detected by The FAST All Sky \HI\ Survey (FASHI) project. Due to the faint and irregular morphology of these galaxies, the default photometry is often inaccurate. Therefore, we utilized The Dark Energy Camera Legacy Survey (DECaLS) data to perform careful photometric measurements, and find that the low \HI\ mass galaxies have similar stellar mass densities to dwarf elliptical (dE) galaxies. Compared to other dwarf galaxy populations, the \HI-selected dwarfs exhibit higher stellar mass densities than ultradiffuse galaxies, and similar densities to \HI-selected low-surface-brightness galaxies, albeit with lower stellar masses, suggesting a possible evolutionary connection among these populations. By classifying the galaxies according to their \HI\ spectral-line profiles, we show that the double-peaked sources conform closely to the Tully–Fisher relation, whereas the single-peaked sources follow the Faber-Jackson relation but with large scatter. This indicates that the single-peaked systems are likely dispersion dominated and that the relationship between stellar mass and halo mass in such systems may remain consistent across both low- and high-mass regimes. These findings suggest that \HI-selected dwarf galaxies with single-peaked \HI\ profiles may share a similar dynamical state with massive ellipticals, offering new insights into their structural evolution and the diversity of formation pathways for low-mass galaxies.
\end{abstract}



\section{Introduction}
\label{sec:intro}

Low-mass (dwarf) galaxies are the building blocks of the hierarchical galaxy formation scenario and provide critical insights into the emergence of fundamental galaxy properties such as mass, metallicity, and dynamics. Studies of the low-mass galaxy population are essential for addressing when and how massive galaxies establish their scaling relations \citep[e.g.,][and references therein]{2008Natur.455.1082D, 2021FrASS...8..157D, 2023A&A...675A.186D}. These galaxies may harbor intermediate-mass black holes in their nuclei \citep{2019NatAs...3..755W, 2024arXiv240511750Z}, offering unique windows into the early co-evolution of black holes and their hosts \citep{2020ARA&A..58..257G, 2013ARA&A..51..511K, 2022NatAs...6...26R}.

Local low-mass galaxies exhibit extreme physical properties \citep{2019ARA&A..57..375S}, serving as analogs for high-redshift galaxies \citep{2023MNRAS.525.2087B}. Their study advances our understanding of star formation in metal-poor environments \citep{2016ApJ...822...62B}, Ly$\alpha$ photon escape during reionization \citep{2021ApJ...923L..28Y}, and galactic merger processes \citep{2008ApJ...677...37O, 2023MNRAS.524.2224L}. The faint-end stellar mass function further constrains stellar feedback mechanisms \citep{1997ApJ...481..703S, 2011IAUS..277..273S}.

Cosmologically, low-mass galaxies predominantly reside in the subhalos of massive host halos \citep{2008Natur.454..735D} and along dark matter filaments tracing the cosmic web. This multifaceted importance establishes low-mass galaxies as crucial laboratories for probing galaxy evolution physics.

However, studies of low-mass galaxies remain limited by insufficient redshift measurements. The intrinsic faintness of dwarf galaxies introduces significant uncertainties in optical photometry, making photometric redshift determination particularly challenging \citep[e.g.,][]{2021ApJS..256....4C}. Spectroscopic redshifts face similar limitations due to instrumental sensitivity thresholds. For instance, a galaxy with stellar mass $10^8 M_\odot$ at $z=0.1$ typically exhibits an $r$-band magnitude of $\sim$21 AB \citep{2018MNRAS.475..788M}, fainter than the detection limits of major wide-field spectroscopic surveys like the Sloan Digital Sky Survey \citep[SDSS;][]{2000AJ....120.1579Y, 2002AJ....124.1810S, 2006AJ....131.2332G}, Galaxy and Mass Assembly \citep[GAMA;][]{2011MNRAS.413..971D, 2015MNRAS.452.2087L, 2018MNRAS.474.3875B, 2022MNRAS.513..439D},  The Large Sky Area Multi-Object Fiber Spectroscopic Telescope survey\citep{2012RAA....12.1197C, 2019RAA....19..113Z} and Dark Energy Spectroscopic Instrument \citep[DESI][]{2016arXiv161100036D, 2024AJ....168...58D}. The ongoing and upcoming wide field spectroscopic redshift survey projects such as Deep Extragalactic Visible Legacy Survey \citep[DEVILS][]{2018MNRAS.480..768D}, Hobby-Eberly Telescope Dark Energy Experiment \citep[HETDEX][]{2021AJ....162..298H, 2023ApJ...943..177M}, 4-metre Multi-Object Spectroscopic Telescope \citep[4MOST][]{2019Msngr.175....3D}, Subaru Prime Focus Spectrograph \citep[PFS][]{2016SPIE.9908E..1MT}, Multi-Object Optical and Near-infrared Spectrograph \citep[MOONS][]{2020Msngr.180...10C}, MUltiplexed Survey Telescope \citep[MUST][]{Li:24, Zhang:24} will be able to reach a deeper magnitude limit (e.g., $m_r \sim 22$ AB mag). Moreover, the redshift identification of the faint galaxies are biased to the strong emission line galaxies such as blue compact dwarf \citep{1981ApJ...247..823T, 2000A&ARv..10....1K}, green pea \citep{2009MNRAS.399.1191C}, blue berries \citep{2017ApJ...847...38Y}, emission-line dots \citep[ELdots,][]{2015MNRAS.454L..41B}, or with a more general name like extreme emission-line galaxies \citep[EELG][]{2024arXiv240417415D}, while the faint galaxies with weak emission line such as UDGs \citep{2015ApJ...798L..45V}, or red compact dwarf galaxies \citep[e.g.,][]{2003Natur.423..519D, 2020ApJS..250...17L}, which usually have absorption lines for redshift identification, and thus out of the spectroscopic redshift detection limit. 

A promising approach to circumvent redshift limitations involves studying dwarf galaxies within the Local Group or galaxy clusters \citep[e.g.,][]{1985ApJ...295...73K, 2004ApJS..153..223C, 2009ApJS..182..216K, 2012ApJS..198....2K, 2012ApJS..200....4F, 2016ApJ...817...84K}. Existing dwarf galaxy research predominantly focuses on Local Group members \citep[e.g.,][]{2019ARA&A..57..375S}, where proximity enables precise distance measurements, or on satellite systems near massive galaxies \citep{2017ApJ...847....4G, 2024arXiv240414499G} that share redshifts with their primaries. However, such satellite populations likely experience evolution dominated by dense environmental effects \citep{2010ApJ...721..193P}, potentially obscuring their intrinsic properties. Consequently, the current census of faint galaxies with robust redshift determinations remains critically incomplete, particularly for isolated field populations.

\HI\ blind surveys conducted with telescopes offer a unique window into galaxy evolution by characterizing neutral hydrogen properties inaccessible to optical studies. The \HI\ 21cm emission line provides two critical diagnostics: (1) redshift determination independent of stellar light, particularly crucial for studying faint systems \citep[e.g., ][]{1998A&AS..127..409K, 2000A&AS..146..359K, 2001A&A...377..801H, 2016A&A...587L...3C, 2019MNRAS.490.3365N, 2021A&A...646A..35M, 2021ApJ...915...70W, 2022A&A...657A..56K, 2023ApJ...943...93M, 2023A&A...673A.146S, 2024A&A...688A.109D, 2025arXiv250515983S, 2025arXiv250519248N}; (2) gas-mass estimates through the \HI\ emission line, with both being vital for studying the gas abundance in galaxies. The blind \HI\ survey proves indispensable for low-surface-brightness galaxies \citep{2015AJ....149..199D, 2019MNRAS.483.1754D, 2019ApJ...880...30H, 2020ApJS..248...33H, 2020NatAs...4..246G, 2020ApJ...902...39K, 2023ApJ...948...96C, 2023ApJ...947L...9H}; almost optically dark galaxies \citep{2005ApJ...622L..21M, 2015AJ....149...72C, 2015ApJ...801...96J, 2017ApJ...842..133L, 2018AJ....155...65B, 2022ApJ...926..167J, 2024ApJ...964...85D, 2024A&A...681A..15M, 2025arXiv250603678K}; or large-scale \HI\ filaments tracing cosmic web \citep[e.g., ][]{2019AJ....157...81O}.

Moreover, \HI\ emission lines not only provide redshift and \HI\ gas mass, but also reveal the rotation velocity through the width of the line profile. The velocity resolution of modern \HI\ surveys ($\Delta v \lesssim 10$ km s$^{-1}$) enables kinematic modeling of dark matter halos even for galaxies with $M_{\rm HI} < 10^8 M_\odot$ \citep{2020ApJS..250...17L}. The line-profile information is crucial for studies of the (baryonic) Tully-Fisher relation \citep{1977A&A....54..661T, 2000ApJ...533L..99M, 2017AJ....153....6K, 2025arXiv250522727M}, which describes a tight correlation between the rotation velocity and the baryonic mass (i.e., stellar plus gas mass) of galaxies. This tight correlation implies a strong connection between the dark matter halo and the distribution of baryons in galaxies. The baryonic Tully-Fisher relation (BTFR) can be constructed by \HI\ emission lines, which provide redshift, \HI\ gas mass, and line-of-sight rotation velocity. Recent studies have shown that the low-mass end of the BTFR is still underconstrained due to a lack of large samples, and that ultra-diffuse galaxies may deviate from the BTFR followed by more massive systems \citep{2019ApJ...883L..33M, 2024ApJ...964...85D, 2024arXiv240400555R}. These findings highlight the importance of assembling large low-mass galaxy samples and conducting follow-up studies.

\HI\ blind surveys also offer an advantage in identifying low\-mass galaxies. The \HI-to-stellar mass ratio exceeds unity when the stellar mass is below approximately $10^9,M_\odot$ for gas rich galaxies \citep{Huang2012}. As a result, galaxies with low \HI\ masses detected in \HI\ surveys are commonly dwarf galaxies, with redshifts directly obtained from their \HI\ emission. Ongoing imaging surveys such as the Dark Energy Camera Legacy Survey (DECaLS) reach a depth of $\sim$24 AB mag in the $g$ band, allowing the detection of galaxies with surface brightness as faint as $\sim$29 mag arcsec$^{-2}$ \citep{2022MNRAS.515.5335L}. Therefore, the combination of deep optical imaging and \HI\ blind surveys can place strong constraints on galaxy properties, particularly at the low-mass end.

FASHI \citep{2024SCPMA..6719511Z} is (to date) the largest \HI\ blind survey conducted with the Five-hundred-meter Aperture Spherical radio Telescope \citep[FAST;][]{2020RAA....20...64J, 2020Innov...100053Q}. It covers approximately 7600 deg$^2$ with an rms sensitivity of about 0.5 mJy. The first data release (DR1) catalog contains over 40,000 \HI-bright sources within a redshift of $z < 0.08$. The data were processed using the FAST data reduction pipeline \citep{2024SCPMA..6759514J}, followed by visual inspection of the spectra and moment-0 maps. The redshift range covered by FASHI is similar to that of ALFALFA, but with higher sensitivity, allowing for the detection of lower \HI\ mass galaxies at comparable redshifts. In addition, the blind nature of the survey has at least doubled the number of known OH and \HI\ absorption-line galaxies, enabled by its unprecedented sensitivity \citep{2024ApJ...971..131Z, 2025ApJS..276....6Z}.

At the low-mass end, FASHI is capable of detecting \HI\ masses as low as $M_{\rm HI} \sim 10^7\,M_\odot$, allowing for the construction of a large sample of low-mass galaxies out to distances of 200 Mpc \citep{2024SCPMA..6719511Z}. More importantly, the FASHI project is also expected to identify a substantial number of optically dark (or extremely faint) galaxies \citep[e.g.,][]{2023ApJ...944L..40X}, and even starless dark matter halos that exhibit only \HI\ emission \citep[e.g., Cloud-9 in][]{2023ApJ...952..130Z}, which has been further confirmed by observations with the Karl G. Jansky Very Large Array (VLA; \citealt{2024arXiv240618643B}). These systems—characterized by very low stellar mass or no stars at all—can serve as critical tests of the Lambda cold dark matter ($\Lambda$CDM) model on small scales, and are therefore of fundamental importance to theories of cosmic structure formation.

As a first step toward investigating the optically faint sources in FASHI, it is essential to characterize the general properties of low \HI\ mass galaxies that possess clear optical counterparts. While previous studies based on the ALFALFA survey have provided valuable insights into low \HI\ mass galaxies \citep[$M_*<10^{7.7}M_\odot$][]{2012AJ....143..133H}, the FASHI survey, with its superior sensitivity and wide area, offers a new opportunity to extend the detectable \HI\ mass to a lower level, allowing us to explore a population of extremely low \HI\ mass systems.

In this study, we focus on a sample of dwarf galaxies with $M_{\rm HI} < 10^8\,M_\odot$, selected from the FASHI DR1 catalog. We specifically aim to investigate the galaxies with the lowest detected \HI\ masses in FASHI, examining their optical surface brightness characteristics and \HI\ spectral profiles. This \HI-based definition is not complete in stellar mass and misses gas-poor dwarf galaxies, but it effectively highlights the low \HI\ mass dwarf galaxies detected by FASHI.

We note that the adopted low \HI\ mass cut ($M_{\rm HI} = 10^8\,M_\odot$) may introduce a selection bias against low-stellar mass but \HI-rich system. At the lowest halo or baryonic masses, galaxies are theoretically expected to be dispersion dominated regardless of their gas content, owing to their shallow potential wells and inefficient angular momentum support \citep[e.g., ][]{2024MNRAS.532.2558Z}. This potential bias should be kept in mind when interpreting the dynamical classifications. Nevertheless, our goal here is to characterize the lowest \HI-mass galaxies detected by FASHI, and to explore how their structural and dynamical properties compare with other low-mass galaxy populations.

The $M_{\rm HI} = 10^8\,M_\odot$ threshold approximately corresponds to a halo mass of $\sim10^{10}\,M_\odot$ \citep[based on typical $M_\star$–$M_{\rm halo}$ relations in e.g., ][]{2021NatAs...5.1069C}. This value is close to the critical halo mass of $5\times10^{9.5}\,M_\odot$, below which dark matter halos are theoretically predicted to be inefficient at forming stars and unlikely to host luminous galaxies in the present day \citep{2017MNRAS.465.3913B, 2020MNRAS.498.4887B, 2023MNRAS.524.2290N}.

By combining DECaLS images in the $g$ and $r$ bands, we can identify the optical counterparts of \HI-selected low-mass galaxies and estimate their basic properties. The deep optical imaging also aids in morphological classification and provides more reliable axis ratio ($b/a$) measurements, which are essential for correcting the \HI\ line width to obtain rotation velocities. However, at redshifts $z < 0.01$, the angular sizes of low-mass galaxies become large and clumpy, and automated photometric pipelines may fragment these galaxies into multiple segments. In addition, contamination from foreground stars in the Milky Way or background bright galaxies can affect total flux measurements. The large beam size of single-dish radio telescopes further complicates the association between \HI\ sources and their true optical counterparts. A statistically large sample and scaling relations can help reduce misidentifications. Meanwhile, more refined photometric segmentation methods are necessary to accurately measure faint, diffuse, or irregularly shaped galaxies. Therefore, we begin with the FASHI DR1 catalog and DECaLS imaging, and conduct careful photometric analysis to investigate the properties of low-mass galaxies.

Throughout this paper, we assume a standard $\Lambda$CDM cosmology with $H_0=70\, \rm km/s/Mpc$, $\Omega_{M} = 0.3$, and $\Omega_{\rm \Lambda} = 0.7$. All the magnitudes are in the AB magnitude system \citep{1983ApJ...266..713O}.

\section{Sample selection}

\subsection{\HI\ Low-mass Galaxy Sample and Optical Counterparts Identification}

We select sources with $M_{\rm HI} < 10^8 M_\odot$ from the FASHI DR1 catalog and crossmatch their coordinates with the DECaLS DR9 footprint.\footnote{Although DECaLS DR10 was released during the completion of this work—providing extended area coverage and additional $i$-band data—we retain DR9 data in this analysis for consistency. FASHI is still ongoing, and a future data release will include a larger sample, with crossmatched results from both DECaLS and DESI.} This crossmatching yields 520 targets with available $g$- and $r$-band imaging. 

We download the DECaLS $g$- and $r$-band images centered on the FAST coordinates. Since many \HI\ targets lie near the edges of the DECaLS image tiles, we first identify the corresponding tile IDs, and then retrieve the science frames (with suffixes {\sc `-image-g.fits.fz'} and {\sc `-image-r.fits.fz'}) along with their inverse variance maps ({\sc `-invvar-g.fits.fz'} and {\sc `-invvar-r.fits.fz'}). The images are mosaicked using {\sc SWARP}\footnote{\url{https://github.com/astromatic/swarp}} and cut into $5\times5$ arcmin$^2$ stamps for further analysis.

We visually identify optical counterparts for the \HI-selected targets. Since low-mass gas-rich galaxies with $M_* < 10^9 M_\odot$ typically have $M_{\rm HI} \gtrsim M_*$ \citep{2012ApJS..200....4F}, their optical counterparts are expected to be relatively faint. The FAST beam size is approximately $3'$, which can cover multiple optical galaxies in the nearby Universe. For the local dwarf galaxy regime studied here, an \HI\ mass of $10^8\,M_\odot$ corresponds to a diameter of $\sim 5$ kpc (or $\sim 48''$ at a distance of 20 Mpc), which is still below the FAST beam size. The astrometric offset between FASHI \HI\ positions and optical sources is typically less than 1 arcmin \citep{2024SCPMA..6719511Z}. The typical \HI\ size and the astrometric accuracy allow us to search for optical counterparts within about 1' around the \HI\ centroid in the optical images.

In most cases, we identify a single faint, blue galaxy near the center of the beam (distance $< 1'$), which is treated as the likely optical counterpart. We show the \HI\ spectra and the g-r-z color image from DECaLS in Figure \ref{SPHI} and \ref{DPHI}, and in online version of this work. This identification will be further validated in the next steps. 

In addition to systems with a clear single counterpart, we identify three categories of special cases: 
(1) {\it \HI\ tidal features or associated \HI\ clouds}: nine low-HI-mass targets\footnote{IDs: 61084, 10696, 12526, 20974, 21490, 21521, 32964, 36251, 36267} appear to trace \HI\ extensions of nearby massive galaxies such as NGC 4258 \citep{2021ApJ...922L..21Z}, or \HI\ clouds surrounding galaxies \citep[e.g.,][]{2023ApJ...952..130Z}.
(2) {\it Multiple dwarf galaxies within the beam}: eight targets\footnote{IDs: 4778, 9771, 19877, 24751, 26423, 29664, 31103, 57621} contain two nearby dwarf galaxies within the FAST beam, potentially indicating ongoing dwarf-dwarf mergers \citep[e.g.,][]{2020ApJ...900..152Z}. The measured \HI\ flux thus likely represents the combined emission.
(3) {\it Low-HI-mass early-type massive galaxies}: a few elliptical or S0 galaxies, such as NGC 315 and UGC 05745, also appear in the low-\HI-mass regime. These systems are known to host low amounts of \HI\ and will be analyzed in a separate, dedicated study. They are excluded from this work.
In addition, we remove several \HI\ sources for which no clear optical counterparts are detected in DECaLS imaging, as they fall beyond the scope of this paper.

\begin{figure}[ht!]
    \centering
    \includegraphics[width=0.5\textwidth]{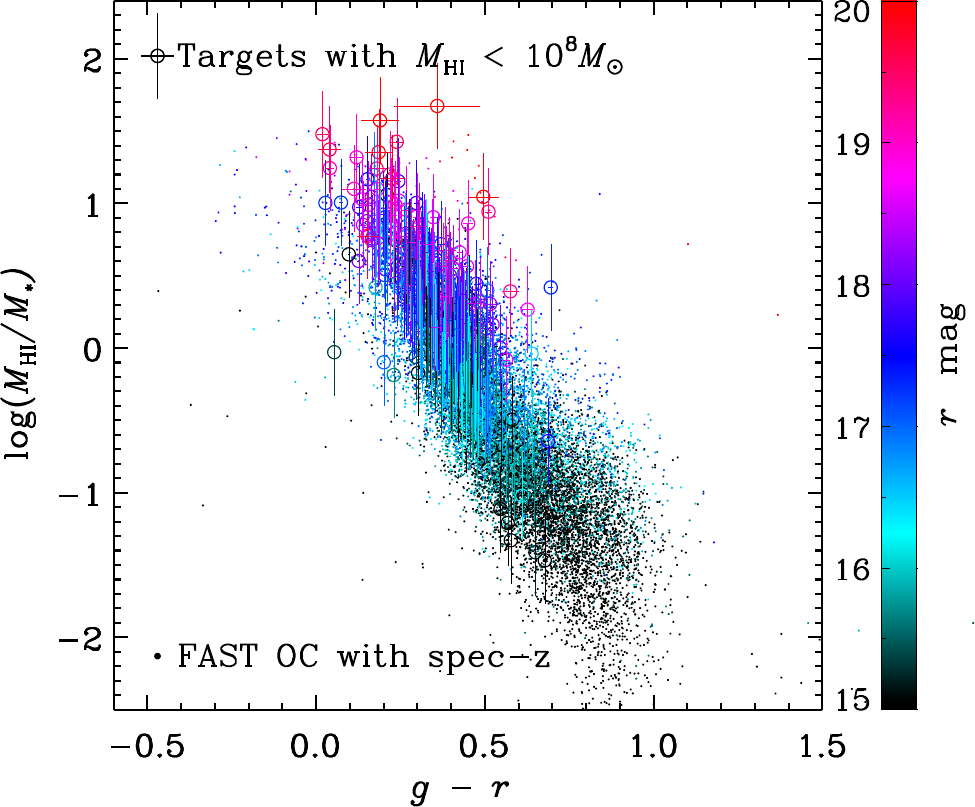}
    \caption{The \HI-to-stellar-mass ratio as a function of $g - r$ color for the low \HI\ mass sample presented in this work (open circles), overplotted with FASHI targets that have reliable optical counterparts (filled circles) as a reference. All points are color coded by their $r$-band magnitude. The similar trend between the two samples indicates the statistical reliability of the optical counterpart identification for our \HI-selected galaxies.
    }
    \label{MassRatio}
\end{figure}

\begin{figure}
    \centering
    \includegraphics[width=0.95\linewidth]{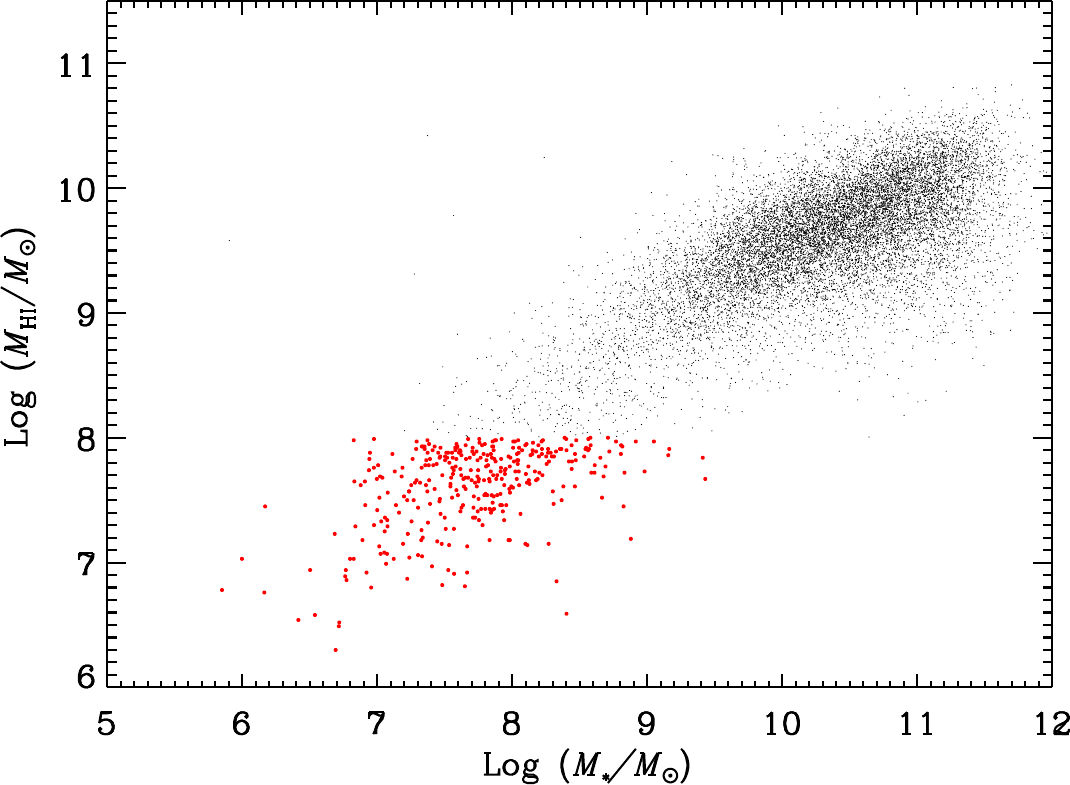}
    \caption{Stellar mass versus \HI\ mass for the FASHI sample, with optical counterparts matched from the SGA catalog with spectroscopic redshifts. The \HI-selected dwarf studied in this work is highlighted in red.}
    \label{MstarMHI}
\end{figure}

\begin{figure*}[ht!]
    \centering
    \includegraphics[width = 0.9\linewidth]{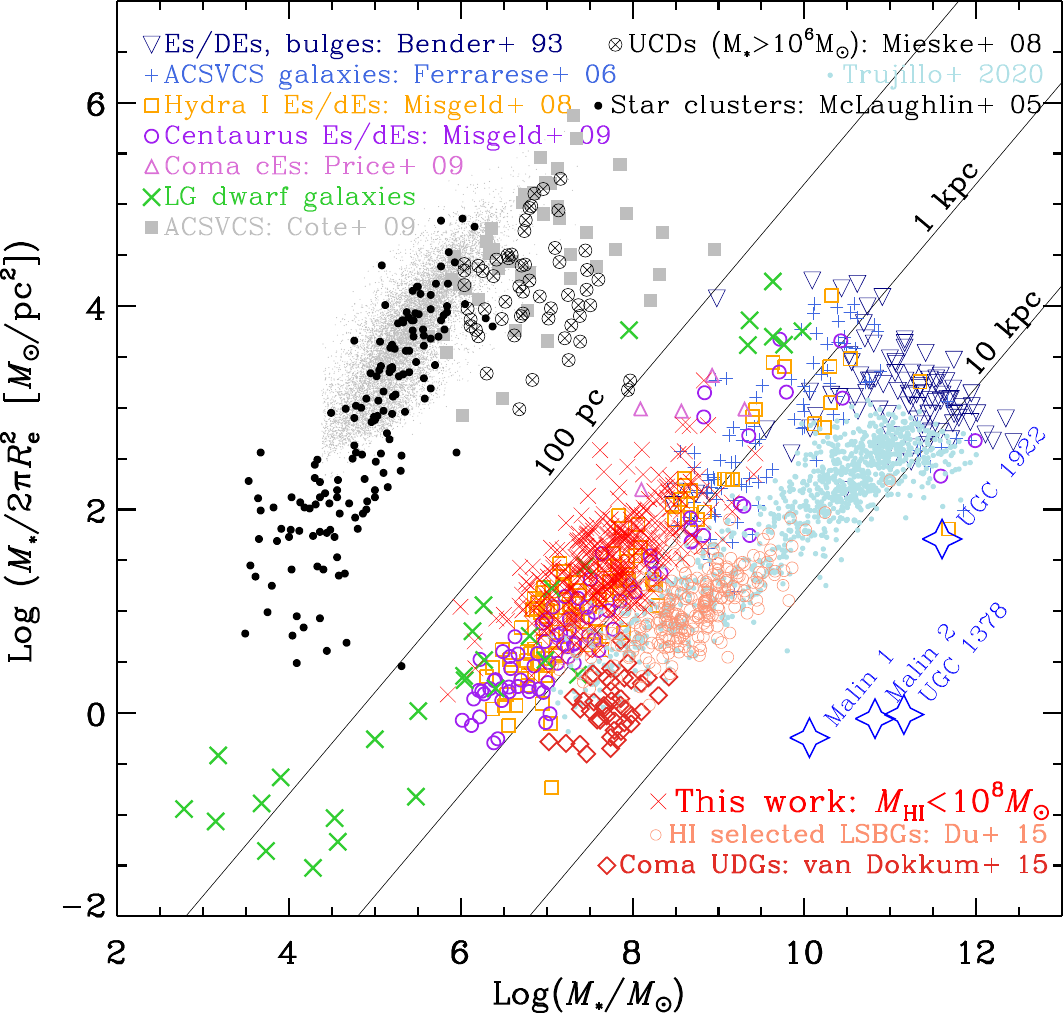}
    \caption{Stellar mass and mass surface density of various stellar systems, including 
    star forming galaxies \citep[light blue dots and blue plus,][]{2006ApJS..164..334F, 2020MNRAS.493...87T},
    elliptical galaxies and early-type galaxies \citep[open downward triangles and gray squares][]{2006ApJS..164..334F, 2006ApJS..165...57C}. 
    To compare with the dwarf galaxy population, we also show the 
    Local Group dwarf galaxies \citep[green crosses,][]{2011MNRAS.414.3699M}, 
    ultracompact dwarfs \citep[black open circles with crosses,][]{2008A&A...487..921M}, 
    dwarf elliptical galaxies \citep[dEs; orange squares and open purple circles, ][]{2008A&A...486..697M, 2009A&A...496..683M, 2009AJ....138.1037C}, 
    low-surface-brightness galaxies \citep[light salmon,][]{Du2015}, and UDGs from coma \citep[dark red,][]{2015ApJ...813...23V}. 
    The low \HI\ mass galaxies are shown with red cross signs, which are at the similar region as dwarf elliptical galaxies. 
    The surface density is roughly 1-2 orders of magnitude lower than the mass density of elliptical or UCDs, implying that no bright bulge formed when the dwarf galaxies were rich in \HI\ gas. 
    We also show the star cluster systems in gray and black dots \citep{2005ApJS..161..304M, 2009ApJS..180...54J}.
    The solid lines indicate the galaxy size of 100 pc, 1 kpc, and 10 kpc.
    }
    \label{massdensity}
\end{figure*}

\begin{figure*}
    \centering
    \includegraphics[width=0.32\linewidth]{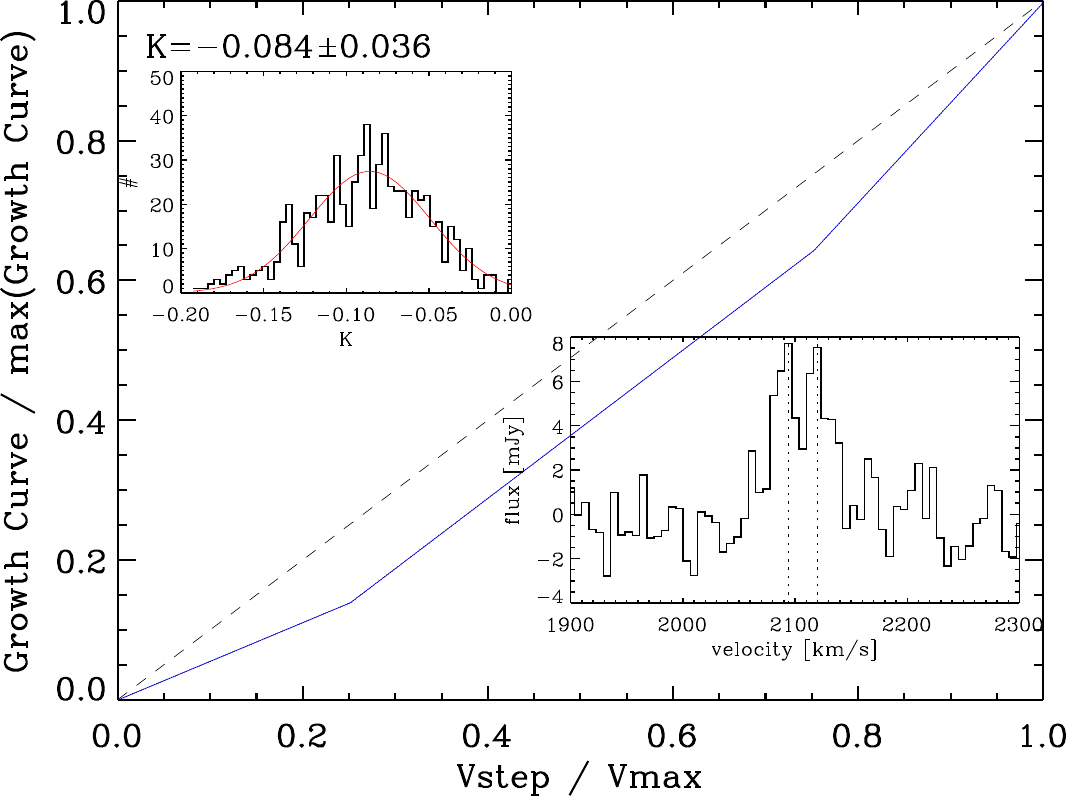}
    \includegraphics[width=0.32\linewidth]{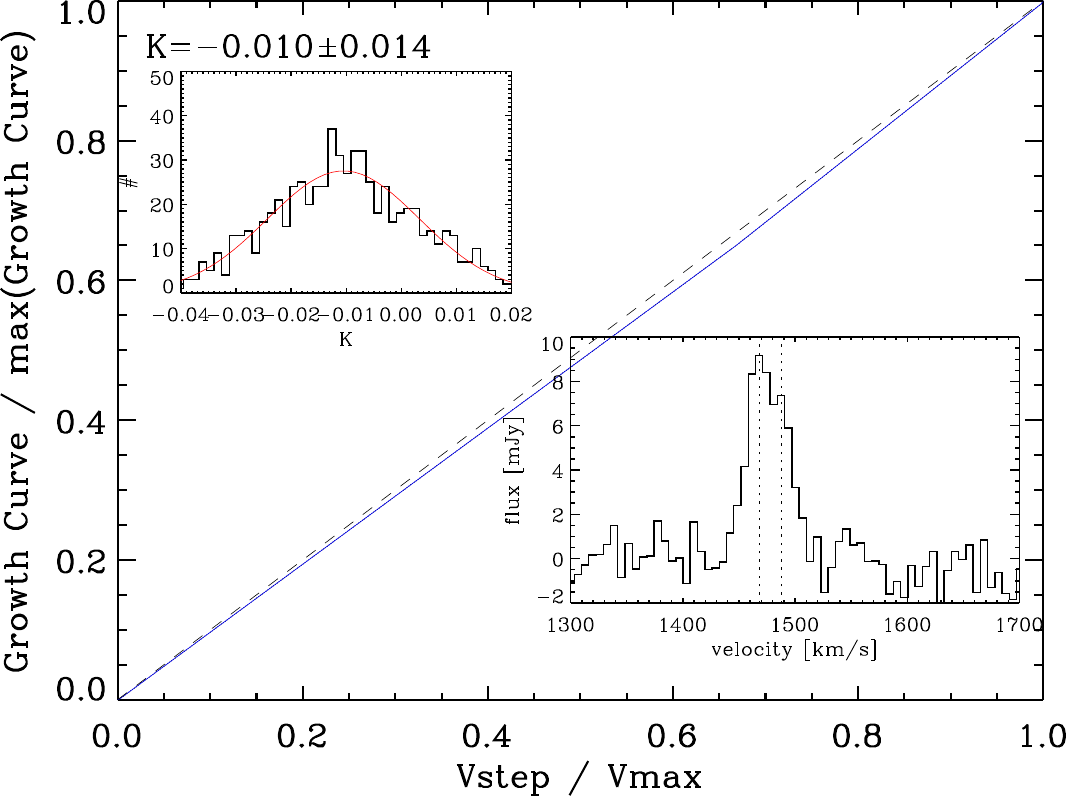}
    \includegraphics[width=0.32\linewidth]{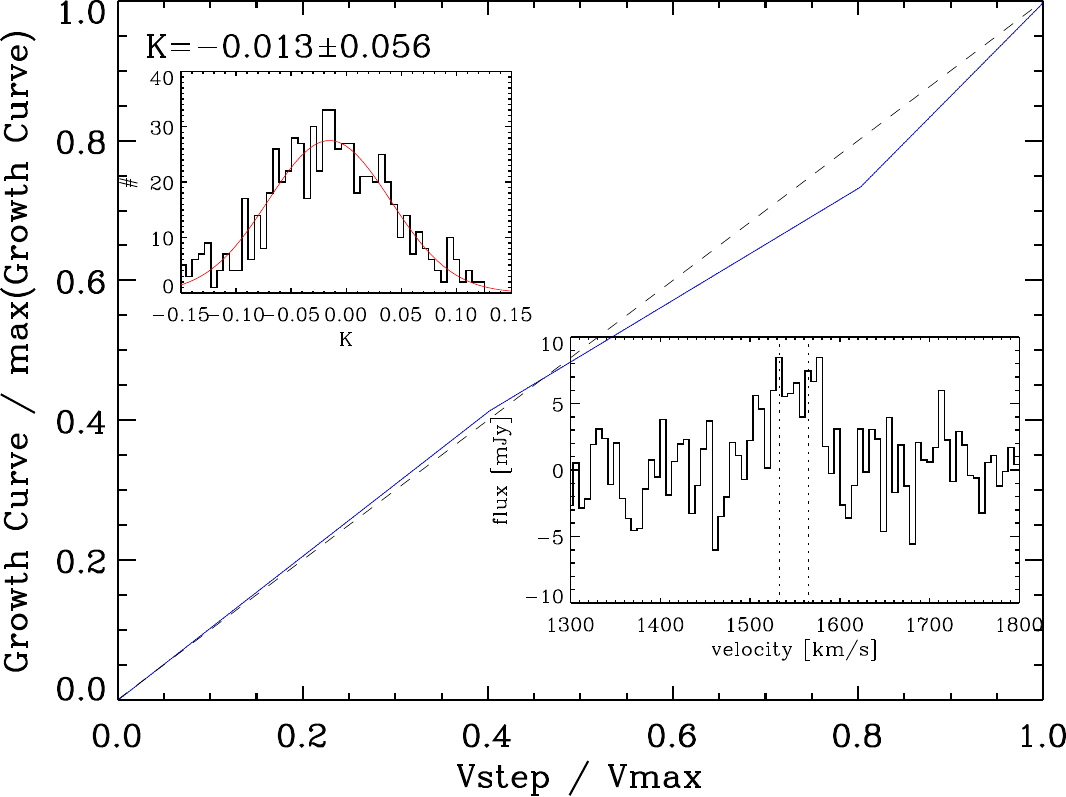}
    \caption{Examples of the growth curve for three targets (ID from left: 40703, 35982, 40772). The left and right peaks are indicated by dashed lines in the lower-right panel of each subplot. The apparent dip between the two peaks is likely caused by noise fluctuations, or not significant enough for the S/N. To account for this, we add random noise at the rms level to each channel 500 times and remeasure the K parameter. The resulting distribution of K values is shown as a histogram in the upper-left panel of each subplot. In the left example, the condition K + Kerr $<$ 0 is satisfied, and the spectrum is classified as double peaked. The right panel example also exhibits a double-peaked profile by eye, but the significance is insufficient for a robust classification in this work.}
    \label{growthcurve}
\end{figure*}

\subsection{Photometry of the dwarf galaxies}\label{photometry}

Approximately 27\% of \HI-rich dwarf galaxies exhibit irregular morphologies \citep[e.g., ][]{2022A&A...659A..14P}, which can confuse the performance of standard source extraction and photometric pipelines. In some cases, the star-forming regions in these gas-rich dwarf galaxies are spatially scattered, causing the pipeline to misidentify them as multiple distinct sources in the DECaLS DR9 photometric catalog.

To mitigate this issue, we first convolve the $r$-band images with a $5^{\prime\prime}$ Gaussian kernel to smooth out clumpy substructures. We then perform photometry using SExtractor\footnote{\url{https://github.com/astromatic/sextractor}} in dual-image mode, using the smoothed image for source detection and the original DECaLS $r$- or $g$-band images for photometric measurements. This convolution helps merge discrete clumps into a single detection and suppresses local rms noise, resulting in a photometric KRON aperture that is generally larger than the default aperture. We adopt the {\sc mag\_auto} value as the total magnitude of each galaxy.

We also manually inspect each segmentation map and remove 50 targets that suffer from significant blending with nearby galaxies or bright foreground stars. Since a flux-complete sample is not essential for our study of low-mass galaxies, we choose to exclude these blended sources and instead focus on isolated, well-separated dwarf galaxies. The final photometric sample comprises 351 galaxies. The photometry results are listed in Table \ref{table1}. 

To validate our identification of optical counterparts, we examine the scaling relation between $g - r$ color and the \HI-to-stellar mass ratio ($M_{\rm HI}/M_*$) \citep{2009MNRAS.397.1243Z}. We first use the FASHI sample with known counterparts to define the typical behavior of \HI-rich galaxies in color and mass ratio space to establish a reference trend for this relation. We crossmatch the FASHI and Siena Galaxy Atlas 2020 \citep[SGA][]{2023ApJS..269....3M} catalogs using a spherical separation of $<3'$ and a velocity difference of $|cz_{\rm specz}^{\rm opt} - V_{\rm HI}^{\rm FASHI}| < 500\,{\rm km\,s^{-1}}$, where $z_{\rm specz}^{\rm opt}$ is the spectroscopic redshift from the SGA catalog, and $V_{\rm HI}^{\rm FASHI}$ is the \HI\ velocity from the FASHI catalog. Given the limited optical data, we estimate stellar masses using the mass-to-light ratio relation $\log(M_{\rm *}/L_r) = -0.306 - 0.15 + 1.097 (g-r)$ from \citet{2003ApJS..149..289B}, where the -0.15 is adopted from the caption of Table 7 in \citet{2003ApJS..149..289B}, and is able to correct the results to Kroupa IMF \citep{1993MNRAS.262..545K}. The typical scatter in stellar mass estimated via the color–mass relation is approximately 0.3 dex, which dominates the uncertainty budget and exceeds the photometric uncertainties. A more detailed comparison of $M_*$ measurements is provided in Appendix~\ref{app:mstar_comparison}. We therefore adopt a conservative uncertainty of 0.3 dex for $M_*$. The results from FASHI and SGA catalogs are shown in Figure \ref{MassRatio}, indicating that the $\log(M_{\rm HI}/M_*) > 0$ targets are typically have $g-r$ color $ < 0.5$.

One challenge in deriving \HI\ masses from the FASHI catalog is distinguishing the cosmological (Hubble) velocity from the peculiar velocity, especially for dwarf galaxies that may be gravitationally influenced by nearby massive systems. For the dwarf galaxy sample in this work, we adopt the distances provided in the FASHI catalog, which are corrected using Cosmicflows-3 but may still be biased by peculiar motions, particularly in low-mass systems. Using the derived $g - r$ color and stellar mass from optical imaging, along with the \HI\ mass from the FASHI catalog, we mark the dwarf galaxy sample as open circles in the $M_{\rm HI}$–$M_*$ distribution shown in Figure~\ref{MassRatio}. These points follow the overall trend of the full FASHI sample, suggesting that our optical counterpart identifications are generally robust. We also show the stellar and \HI\ mass of our sample in Figure \ref{MstarMHI}.

\section{Properties of the \HI-selected dwarf galaxies}

To investigate the properties of the \HI-selected dwarf galaxy sample, we consider their stellar structure, gas kinematics, and environmental context. Specifically, we analyze the stellar mass surface density to assess their evolutionary state, examine the \HI\ line profiles (e.g., single peaked or double peaked) to explore their dynamical features, and study their clustering behavior to understand potential environmental influences. Although the FAST telescope lacks the resolution to spatially resolve these systems, their global properties can still offer valuable insights into their formation and evolution.

\subsection{Optical Surface Brightness}

The surface brightness of galaxies is a key parameter for understanding their evolutionary stages \citep[e.g.,][]{2009ApJS..182..216K}. We estimate the stellar mass surface density of our sample and place them in the broader context of local stellar systems. This includes a wide range of low-mass galaxy populations such as ultracompact dwarfs \citep[UCDs;][]{2008A&A...487..921M}, dwarf elliptical galaxies \citep[dEs;][]{2008A&A...486..697M, 2009A&A...496..683M, 2009AJ....138.1037C}, local group dwarf galaxies \citep[][]{2011MNRAS.414.3699M}, and ultra-diffuse galaxies \citep[UDGs;][]{2015ApJ...813...23V}, as well as more massive galaxies found in both field and galaxy clusters \citep{2006ApJS..164..334F, 2006ApJS..165...57C, 2020MNRAS.493...87T}. A few giant low surface brightness galaxies from recent studies \citep{2023ApJ...959..105D} are also included for comparison, along with star clusters \citep{2005ApJS..161..304M}. 

As shown in Figure~\ref{massdensity}, most \HI-selected dwarf galaxies ($M_{\rm HI}<10^9 M_*$) in our sample have stellar surface densities in the range of $10^{1-2.5} \, M_\odot\, \rm pc^{-2}$, comparable to those of dwarf elliptical galaxies (dEs, also known as early-type dwarf galaxies; $\log (M_*/2\pi R_e^2) \sim 10^{0-2} \, M_\odot\, \rm pc^{-2}$) and the \HI-selected low-surface-brightness galaxies in \citet{2023ApJ...959..105D}. These stellar mass surface density values are lower than those of massive stellar mass galaxies ($\gtrsim 10^{2} \, M_\odot\, \rm pc^{-2}$) reported by \citet{2020MNRAS.493...87T}, but higher than those of ultra-diffuse galaxies \citep{2015ApJ...813...23V}, which typically have stellar mass surface densities of about $1 \, M_\odot\, \rm pc^{-2}$. The stellar masses of the \HI-selected dwarf galaxies are in the range of $10^{6.5 - 9} \, M_\odot$, placing them at the low-mass end of the galaxy population. 

We therefore conclude that \HI-selected dwarf galaxies are predominantly low-surface-density systems compared with more-massive galaxies, and have similar stellar surface densities to dEs, which are characterized by low gas content and weak star formation—essentially miniature versions of massive elliptical galaxies \citep{1994A&ARv...6...67F}. Given the generally low star formation rates of HI-selected dwarf galaxies \citep{2012AJ....143..133H}, the similarity in stellar surface density suggests that these systems may be progenitors of dEs in the field \citep[e.g.,][]{2023ApJS..265...57P}. A detailed comparison of their star formation histories, particularly through optical spectroscopy, will be crucial for understanding the evolutionary connection between these two populations.

Moreover, the low stellar surface densities of \HI-rich dwarfs indicate that most have not yet formed a dominant, compact stellar core such as a bulge. Recent studies suggest that such compact cores in dwarfs may be induced by tidal stripping from nearby massive galaxies \citep{2018ApJ...858...37Z, 2020ApJS..250...17L, 2023Natur.623..296W}. Indeed, bright dEs are frequently nucleated \citep{2023ApJS..265...57P}, implying that the removal or depletion of \HI\ gas may occur prior to bulge formation.

\begin{figure*}[ht!]
    \centering
    \includegraphics[width = 0.95\textwidth]{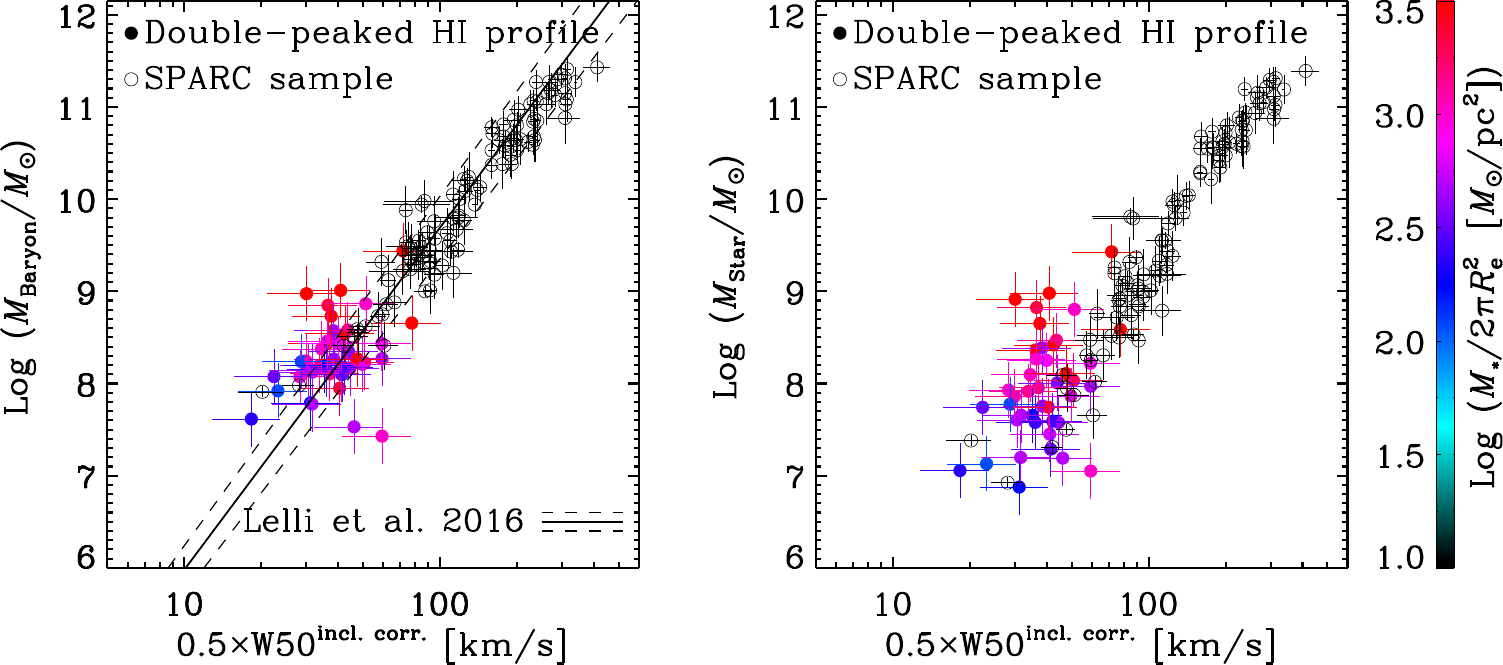}
    \includegraphics[width = 0.95\textwidth]{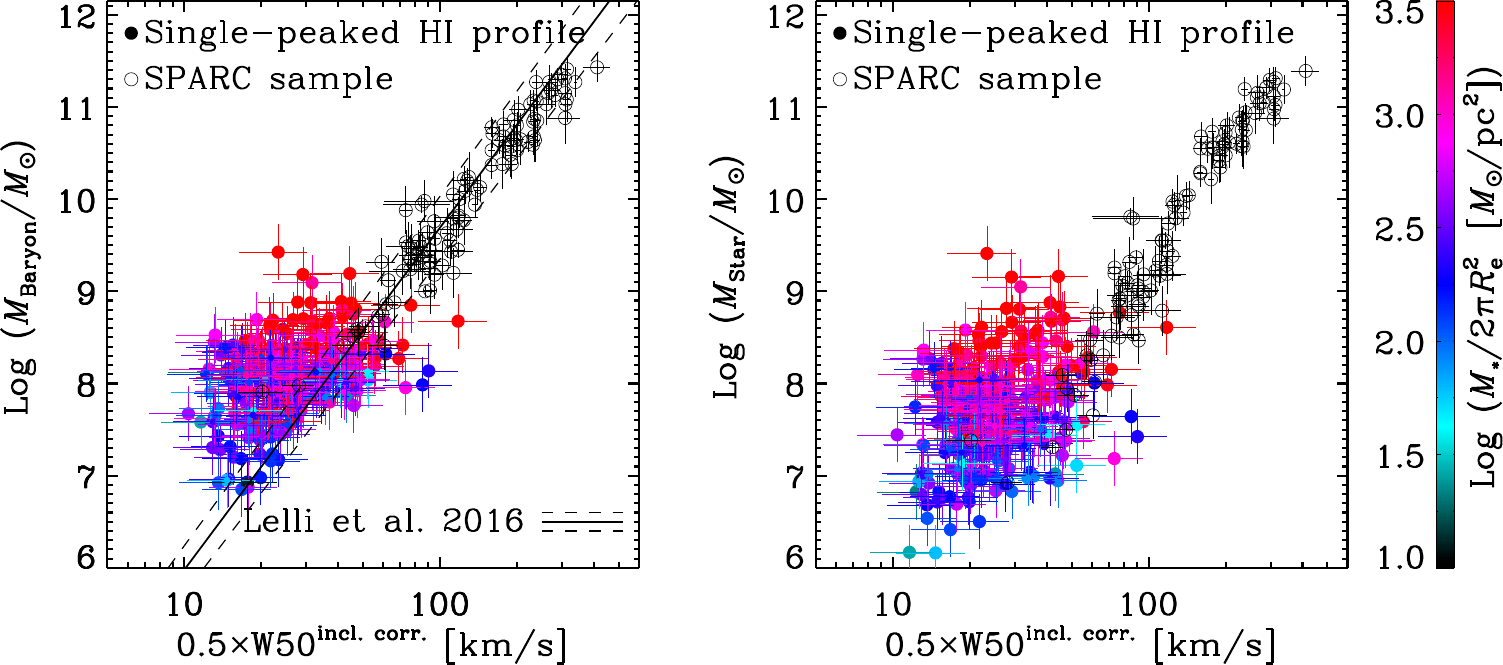}
    \caption{Inclination-corrected $0.5\times$W50 velocity vs the baryonic and stellar mass for double peaked (top panels) and single peaked (bottom panels). The open circles are the sample from the Spitzer Photometry and Accurate Rotation Curves \citep[SPARC, open circles, ][]{2016AJ....152..157L}, which defines baryonic Tully-Fisher relation \citep[black solid line from ][]{2016ApJ...816L..14L}. Our \HI\ low-mass with double-peaked \HI\ profiles follows the trend of the scaling relation, while the single-peaked targets are on or above baryonic Tully-Fisher relation. Our \HI\ sample is colored by the stellar mass surface density.
    }
    \label{TFR}
\end{figure*}

\begin{figure*}
    \centering
    \includegraphics[width=0.49\textwidth]{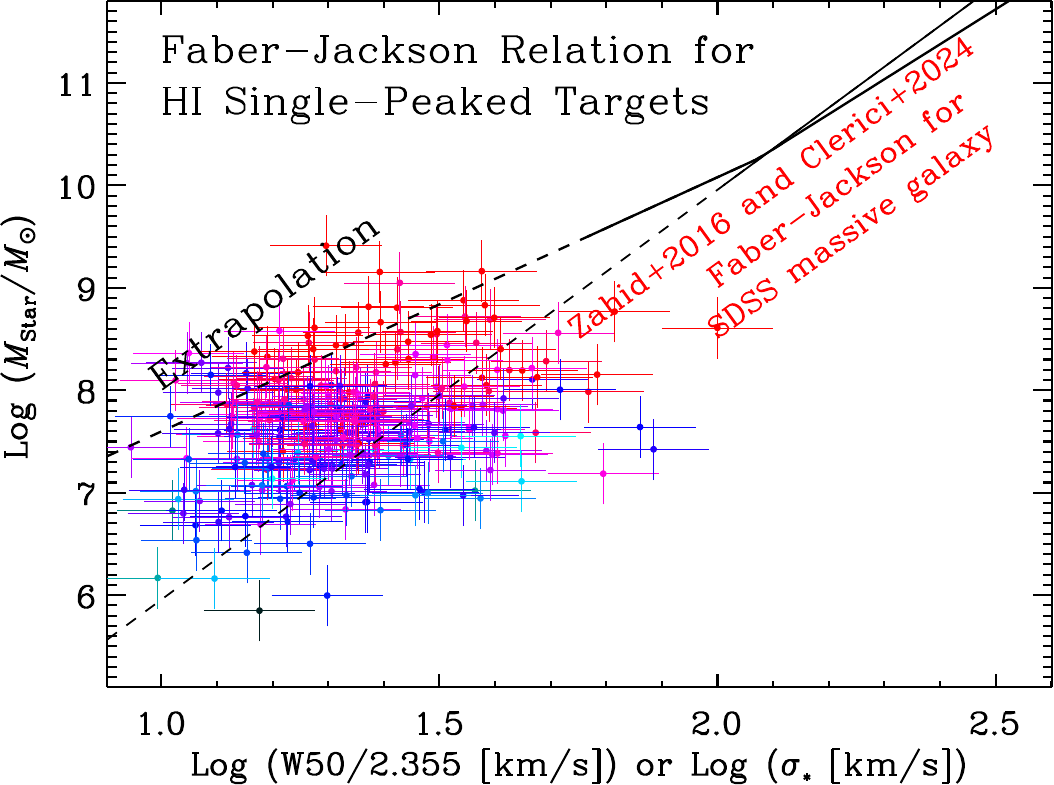}
    \includegraphics[width=0.49\textwidth]{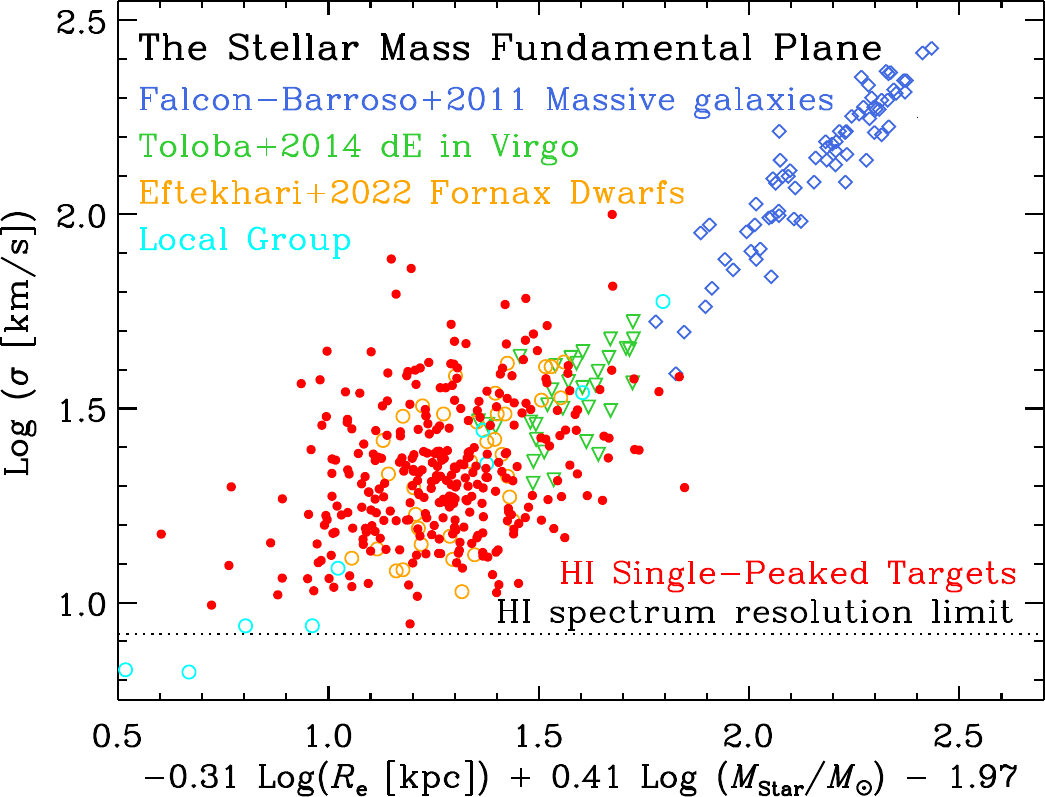}
    \caption{{\bf Left panel:} Faber-Jackson relation for the \HI\ single-peaked targets. We show the results for massive galaxies from \citet{2016ApJ...832..203Z} and \citet{2024MNRAS.531.1034C} in solid lines, and extrapolate to the low-mass region. The \HI\ single-peaked galaxies are shown with dots. We estimate the velocity dispersion with $\sigma = \rm W50/2.355$, and the results are consistent with the trend of the massive galaxies.
    {\bf Right panel:} Fundamental plan for elliptical galaxies sample in \citet[][in blue diamond]{2011MNRAS.417.1787F}; \citet[][in green triangle]{2014ApJS..215...17T}, \citet[][in orange open circles]{2022MNRAS.517.4714E} and galaxies in the Local Group \citep[][in cyan circles]{2006ApJ...642L..37Z}.
    The single-peaked \HI\ targets are shown in red dots, and follow a similar trend and scatter to the dwarf galaxies in Fornax. The dotted line shows the \HI\ emission-line resolution limit of $3\times V_{\rm channel} / 2.355 = 8.3 \rm km\,s^{-1}$. 
    }
    \label{FJFP}
\end{figure*}

\subsection{\HI\ Profile Classification: Single or Double peaked}

Before studying the \HI\ dynamics with FAST spectra, we first classify the \HI\ profiles into single-peaked or double-peaked types, which roughly reflect whether the dynamical state is dominated by velocity dispersion or by rotation. Previous works \citep{2022ApJ...930...85Y, 2022ApJS..261...21Y, 2025arXiv250819545H} have shown that the shape of the growth curve ($F(\Delta v) = \int_{-\Delta v}^{\Delta v} f(v) dv$) indicates whether the \HI\ spectrum has a single peak or double horns. And the K parameter, defined as the integrated offset between the normalized growth curve and the diagonal, is sensitive to the profile shape. Thus, the \HI\ profile can be classified using the K parameter.

On the other hand, the K parameter is not strongly affected by velocity dispersion in bright, \HI-rich galaxies, which typically have steep edges around the two peaks. In such cases, K values measured within either W50 or W20 give consistent results. However, in dwarf galaxies, the velocity dispersion is comparable to the rotation velocity, leading to shallower edges in double-horned \HI\ profiles. Moreover, in the line center, even if the profile appears to exhibit double horns by eye (e.g., two adjacent bright channels separated by a single fainter channel as a dip), the low signal-to-noise (S/N) of these \HI-faint targets makes such features unreliable and introduces significant uncertainties. The S/N of individual \HI\ channels and the spectral resolution are also modest.

Therefore, we modify the classification method. Starting from both sides of W50, we search for the first \HI\ peaks on the left and right sides of the profile. If the two peak velocities are within three channels ($\sim $ 19.2 km/s), we classify the profile as single peaked, even if there is a shallow dip between the peaks. For the remaining targets, we calculate the K parameter between the left and right peak channels: profiles with $K > 0$ are defined as single peaked, while those with $K < 0$ are defined as double peaked. 

For the case of double-peaked profiles, the flux density at the line center is generally weak. To reduce the impact of noise in this region, we added a random noise of the spectrum rms to each channel and repeated the K parameter measurement 500 times. The width of the resulting K distribution is taken as the measurement uncertainty. We then required $K + \sigma_K < 0$ for a spectrum to be reliably classified as double peaked. This error-aware criterion effectively removes most of the ambiguous classifications that could otherwise arise, for instance when a spurious spike in the wing of a single-peaked profile mimics a double peak in noisy data. Among the spectra that classified as single peaked, some of them exhibit flat-topped profiles \citep[e.g.,][]{2022ApJ...930...85Y, 2022ApJS..261...21Y}. As a statistical study to the single- and double-peaked \HI\-selected sample, we do not further distinguish between these single-peaked and flat-topped profiles, and treat both as single peaked, since only high-resolution \HI\ observations can robustly determine the underlying dynamical state. 

Using the above approach, we identify 309 single-peaked targets and 42 double-peaked targets. So most of the dwarf galaxies in our sample have a single-peaked \HI\ profile, which is consistent with the previous results \citep{2020ApJ...902...39K, 2024A&A...684L..24K}. Example spectra and the K parameter measurements are shown in Figure~\ref{growthcurve}.

As can be seen the example spectra in Figures \ref{SPHI} and \ref{DPHI}, some weak sources have very narrow W50 values, for which higher-spectral-resolution and higher-S/N observations are required to accurately assess the line profile.

\subsection{Double-peaked \HI\ dwarf: Tully-Fisher Relations}\label{sec32}

The Tully-Fisher relation is one of the most well-known scaling relations for disk galaxies \citep{1977A&A....54..661T}, and is widely used for distance measurements \citep[e.g.,][]{2016AJ....152...50T, 2023ApJ...944...94T}. This relation connects stellar mass with rotation velocity, reflecting the underlying link between baryonic content and dark matter halo mass. Deep optical long-slit spectroscopy from the DEEP2 Galaxy Redshift Survey \citep{2013ApJS..208....5N} has shown that the Tully-Fisher relation remains tight for massive galaxies \citep{2007ApJ...660L..35K, 2012ApJ...758..106K} out to $z \sim 1$, but becomes increasingly scattered for disklike systems with $M_* < 10^{9.5} \, M_\odot$ \citep{2015MNRAS.452..986S}. In low-mass galaxies, a revised kinematic parameter $S_k = \sqrt{k V_{\rm rot}^2 + \sigma_g^2}$, with $k = 0.5$, shows a better correlation with stellar mass, indicating that these galaxies may still be in the process of disk formation, and that random motions (velocity dispersion) contribute significantly to their gravitational support \citep{2007ApJ...660L..35K, 2012ApJ...758..106K}.

The baryonic Tully-Fisher relation, which links the total baryonic mass to rotation velocity, holds remarkably well over nearly five orders of magnitude in mass \citep{2000ApJ...533L..99M}. However, recent studies suggest deviations at the high-mass end \citep{2019ApJ...884L..11O}, hinting at a more complex or hybrid formation pathway for massive systems. At the low-mass end, several observational works have found that many ultra-diffuse galaxies exhibit significantly lower rotation velocities than expected from their baryonic mass \citep{2019ApJ...883L..33M, 2019MNRAS.484.3267L, 2023ApJ...947L...9H, 2024ApJ...964...85D, 2024arXiv240400555R}, possibly indicating a deficiency in dark matter content.

For \HI\ data without spatially resolved information, rotation velocity is typically estimated from the width of the double-peaked \HI\ profile, corrected by the inclination angle derived from the axis ratio ($b/a$) of optical images. In contrast, low-mass galaxies often exhibit single-peaked \HI\ profiles, suggesting that their line-of-sight velocity is dominated by velocity dispersion rather than rotation. As discussed in Appendix \ref{w50vrot}, we compile a comparison sample of dwarf galaxies with single-peaked \HI\ profiles and high-spatial-resolution \HI\ data from the VLA or Westerbork Synthesis Radio Telescope. The comparison reveals that rotation velocities estimated from inclination-corrected W50 are broadly consistent in trend but systematically offset from those derived directly from the spatially resolved \HI\ data cubes. The inclination-corrected W20 tends to overestimate the true rotation velocity even further (Figure \ref{W50rot}). And in dwarf galaxies with single-peaked \HI\ profiles, W50 may not be a reliable proxy for rotation velocity and instead likely reflects the velocity dispersion of the \HI\ gas. 
So we use W50 to estimate the rotation velocity only for targets with double-peaked \HI\ profiles.

To investigate the evolutionary stage and dark matter content of \HI-selected dwarf galaxies, we present the (baryonic) Tully-Fisher relations for our sample. To mitigate potential biases in rotation velocity estimates from W50 in single-peaked \HI\ systems,  we only consider the double-peaked targets in this section, which are more likely to be the rotation-dominated system. Rotation velocities are derived from W50 using the inclination correction $V_{\rm rot} = \rm W50 / \sin(i)$, where the inclination angle $i$ is obtained from $\cos(i) = \left( \frac{q^2 - q_0^2}{1 - q_0^2} \right)$, following the method adopted in previous studies \citep[e.g.,][]{2019MNRAS.483.1754D}. W50 values are taken from FASHI dr1 catalog \citep{2024SCPMA..6719511Z}. The total baryonic mass is calculated as $M_{\rm baryon} = M_* + 1.3 \, M_{\rm HI}$, including a correction for helium.

The (baryonic) Tully-Fisher relation of our sample are shown in Figure \ref{TFR}. For galaxies with double-peaked \HI\ profiles, our results are consistent with previous (baryonic) Tully-Fisher relations \citep{2000ApJ...533L..99M, 2015MNRAS.452..986S}. In contrast, galaxies with single-peaked \HI\ profiles show a systematic offset toward lower rotation velocities, similar to what has been reported for \HI-rich UDGs \citep{2019ApJ...883L..33M, 2020MNRAS.495.3636M, 2020ApJ...902...39K, 2024arXiv240400555R}, often interpreted as a deficiency in dark matter halo mass. Since our sample has higher stellar mass surface densities than typical UDGs (see Figure \ref{massdensity}), this offset may suggest that the dark matter discrepancy is more closely linked to the distribution and kinematics of the \HI\ gas rather than to optical surface brightness. Nonetheless, high-resolution \HI\ observations remain essential for accurately determining robust rotation velocities, particularly for the single-peaked systems.

\subsection{Single-peak \HI\ dwarf: Velocity-dispersion-dominated System}\label{sec33}

On the other hand, it is also possible that the (baryonic) Tully-Fisher relation does not hold for galaxies with single-peaked \HI\ profiles, as these systems are not rotation-supported like typical spiral galaxies. \HI\ profiles of low-mass galaxies are often single-peaked rather than double-peaked \citep[e.g.,][]{2020ApJ...902...39K, 2024A&A...684L..24K, 2025arXiv250606424J}, indicating a velocity-dispersion-dominated kinematic structure \citep[e.g., dE galaxies;][]{2002AJ....124.3073G, 2003AJ....126.1794G}. The velocity resolution of the FAST data cube is approximately 6.5 km\,s$^{-1}$, which is sufficient to resolve rotation for galaxies at the low-mass end of the (baryonic) Tully-Fisher relation \citep[roughly 30 km\,s$^{-1}$ for a baryonic mass of $10^{6.5} M_\odot$, even for satellite galaxies;][]{2019ARA&A..57..375S}. Therefore, the \HI\ flux in single-peaked galaxies may be confined to a relatively small region, within which the rotation curve has not yet reached the flat part of the galactic rotation velocity. In this case, the observed \HI\ line width does not reflect the true rotational support of the system. Alternatively, the line width may be broadened primarily by turbulent or random motions, i.e., velocity dispersion, rather than ordered rotation, implying a high $\sigma/V_{\rm rot}$. In such systems, the scaling relation between profile width and galaxy mass may be better described by the Faber–Jackson relation \citep{1976ApJ...204..668F}, rather than the Tully–Fisher relation, although both originate from the balance between gravitational potential and internal motion.

To investigate the connection between velocity dispersion and stellar mass, we compare our single-peaked \HI\ sample to massive galaxies ($M_*>10^{10} M_\odot$) from SDSS \citep{2024MNRAS.531.1034C} in the Faber–Jackson relation (left panel of Figure \ref{FJFP}). The single-peaked sample follows an extrapolated trend of the Faber–Jackson relation to lower masses, albeit with a large scatter of about 0.5 dex.

The Faber–Jackson relation is one projection of the fundamental plane of elliptical galaxies, which relates velocity dispersion, surface brightness, and half-light radius \citep{1987ApJ...313...59D, 1987ApJ...313...42D}. Previous studies of dwarf galaxies in galaxy clusters have shown that they deviate from the classical fundamental plane defined by massive galaxies \citep{2005A&A...438..491D, 2012A&A...548A..78T}, possibly due to a varying mass-to-light ratio at the low-mass end. \citet{2022MNRAS.517.4714E} found that this offset is significantly reduced when replacing surface brightness with stellar mass surface density. In the right panel of Figure \ref{FJFP}, we show our \HI-selected sample in the stellar mass fundamental plane for our \HI-selected dwarf sample. The low velocity dispersion dwarfs follow the scaling trend of massive ellipticals and show scatter comparable to that of dwarfs in the Fornax cluster. This consistency suggests that \HI-selected dwarf galaxies may be in a similar dynamical state to massive ellipticals. We conclude that the line width of single-peaked \HI\ profiles traces the velocity dispersion of the gas. Further discussion on velocity dispersion is provided in Section \ref{sigma}.

\begin{figure}
    \centering
    \includegraphics[width=0.48\textwidth]{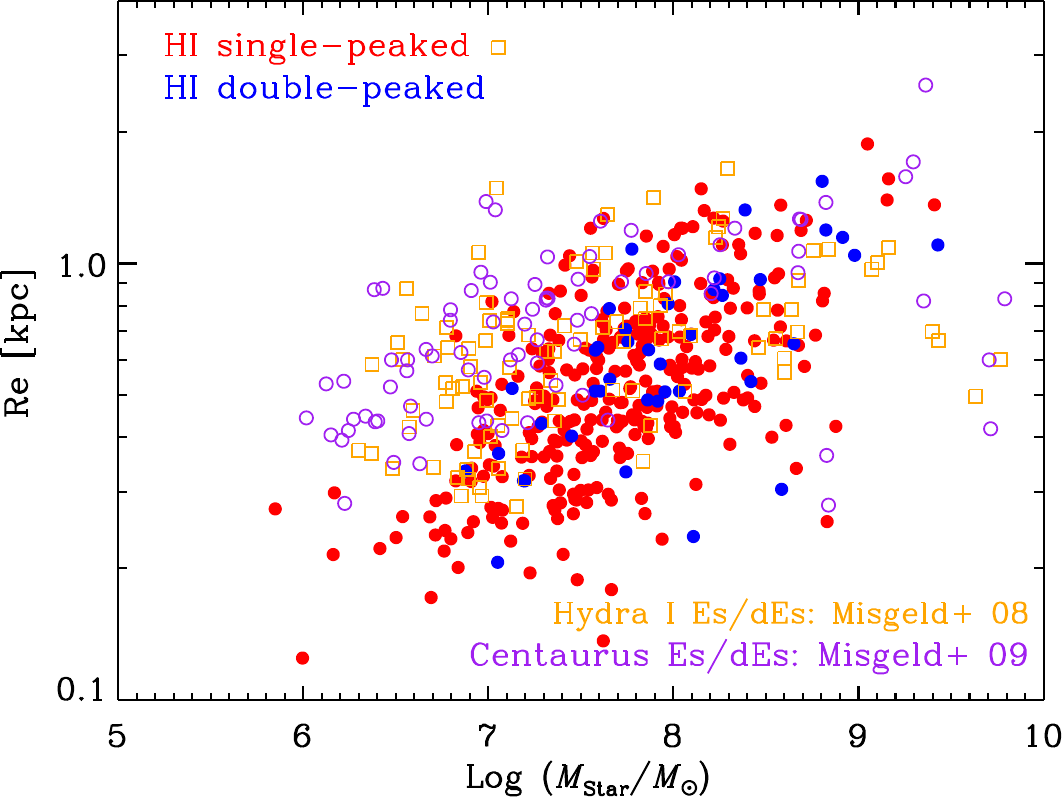}
    \includegraphics[width=0.48\textwidth]{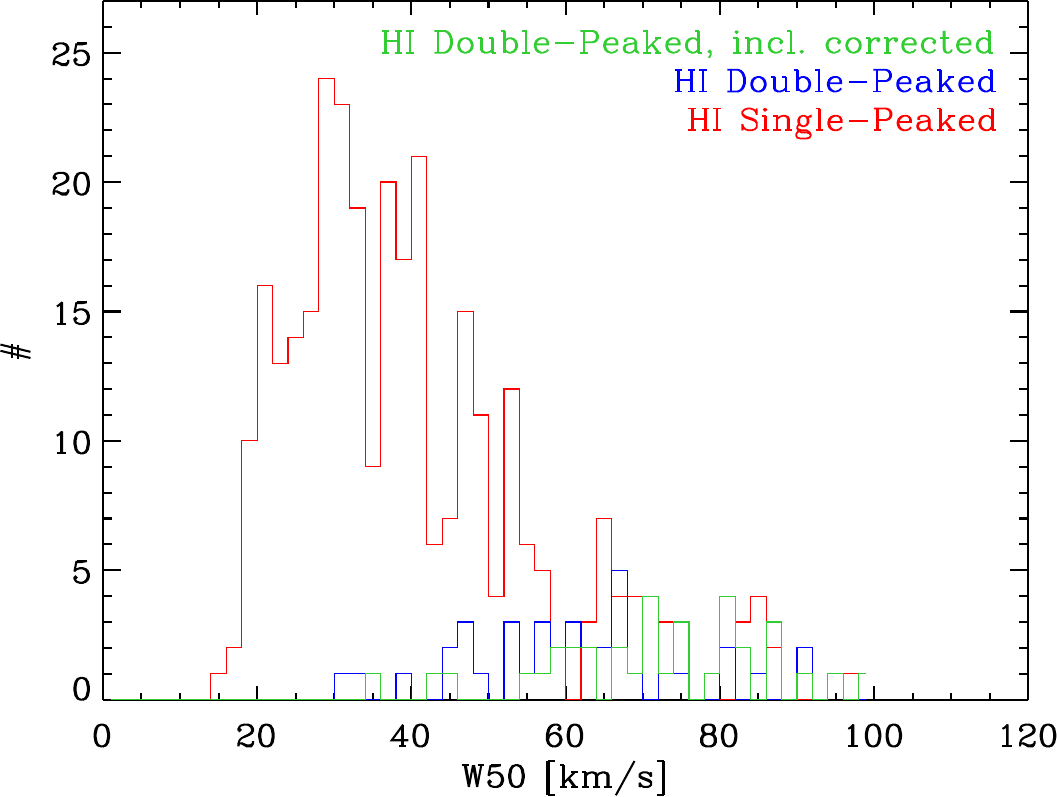}
    \caption{{\bf Top panel:} Stellar-mass size distribution of the low-mass \HI\ sample with single- (red dots) and double-peaked (blue dots) \HI\ profiles. There is no clear offset in the distribution, implying that the stellar size or stellar surface density do not directly connect with the gas dynamics. We also overplot the dE galaxy sample with purple circles and orange squares from \citet{2008A&A...486..697M, 2009A&A...496..683M}, which have a similar distribution as the low \HI\ mass galaxies. {\bf Bottom panel:} Histogram of the W50 for single- and double-peaked \HI\ targets. The single-peaked \HI\ width is systematically lower, which is consistent with a velocity-dispersion-dominated system.
    }
    \label{masssize}
\end{figure}

\section{Discussion}

\subsection{Differences between \HI\ Single- and Double-peaked Galaxies}

We investigate the connection between galaxy properties and the shape of their \HI\ profiles in this section. Galaxies with double-peaked \HI\ profiles generally exhibit higher gas angular momentum \citep{2024arXiv240400555R}, which may also be associated with disklike stellar morphologies. As shown in Figure \ref{masssize}, dwarf galaxies exhibit similar stellar half-light radii regardless of their \HI\ profile types. In contrast, among massive galaxies, those with double-peaked \HI\ profiles are typically spirals, whereas single-peaked \HI\ profiles are more often found in elliptical galaxies, which are dispersion-dominated systems and more compact in size compared to spirals \citep{2012ApJS..203...24V, 2019ApJ...877..103S}. Therefore, for \HI-selected dwarf galaxies, the similarity in stellar sizes despite differences in \HI\ profile morphology suggests a possible decoupling between gas and stellar kinematics. High-resolution observations of both gas and stellar kinematics in dwarf galaxies will be essential for understanding their dynamical states and constraining their formation histories.

\subsection{W50 from Single-peaked \HI\ Profiles and the Rotation Velocity}

The \HI\ line width reflects a combination of projected rotation, velocity dispersion, and other components such as inflow or outflow motions. Although a single-peaked \HI\ profile often implies a more compact \HI\ distribution or velocity-dispersion-dominated system, rather than dominated by ordered rotation, several previous studies still use W50 or W20 as proxies for the rotational velocity ($V_{\rm rot}$) in low-mass galaxies \citep[e.g.,][]{2019MNRAS.484.3267L, 2023ApJ...947L...9H, 2024ApJ...964..135S} when lacking high resolution \HI\ map. For massive galaxies with $V_{\rm rot} \sim 100~\rm km~s^{-1}$ and a typical velocity dispersion of $10\text{–}15~\rm km~s^{-1}$ \citep{2008AJ....136.2782L, 2009AJ....137.4424T}, the line width W50 or W20 can reasonably approximate the rotational velocity. However, in low-mass galaxies with a more compact \HI\ distribution, ordered rotation may still exist but not dominate the dynamics, this assumption becomes less valid.

Our results show that single-peaked \HI\ profiles in dwarf galaxies are more consistent with the Faber-Jackson relation, suggesting that the \HI\ line width is more closely related to velocity dispersion than to rotation. Previous measurements of velocity dispersion are typically derived from stellar absorption lines obtained via medium- or high-resolution optical spectroscopy, which can be time-consuming. For example, \citet{2016ApJ...828L...6V} obtained stellar velocity dispersions using 33.5 hr of integration with Keck/DEIMOS. In comparison, \HI\ spectroscopic observations can achieve a velocity resolution equivalent to $R>10,000$ with significantly less observing time. A deep optical spectroscopic survey targeting a sample of single-peaked \HI\ galaxies would be instrumental in directly comparing stellar and gas kinematics in dwarf galaxies and further assessing the interpretation of W50 as a tracer of dispersion rather than rotation.

\begin{figure}
    \centering
    \includegraphics[width=0.48\textwidth]{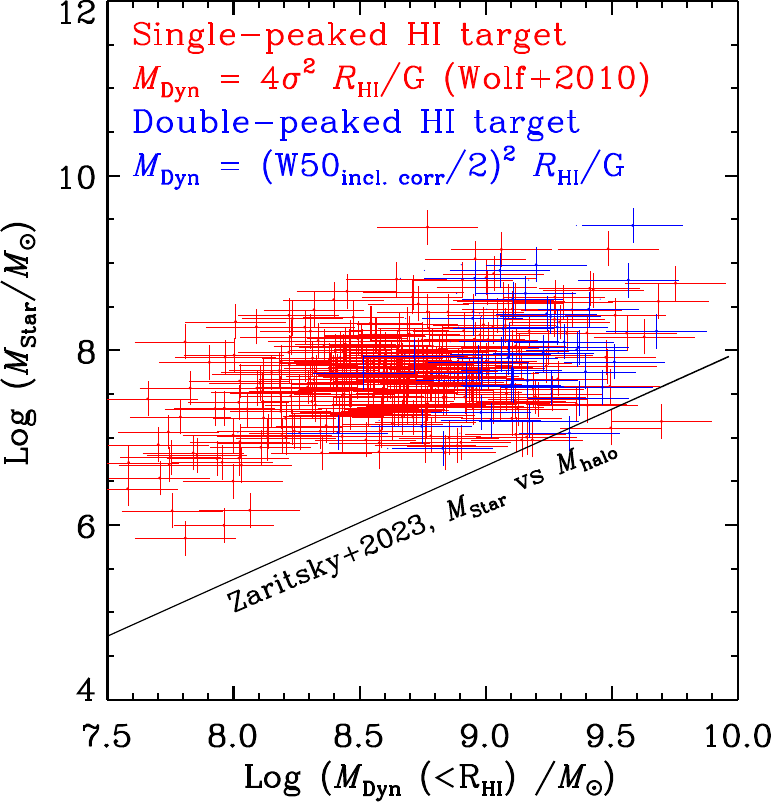}
    \includegraphics[width=0.48\textwidth]{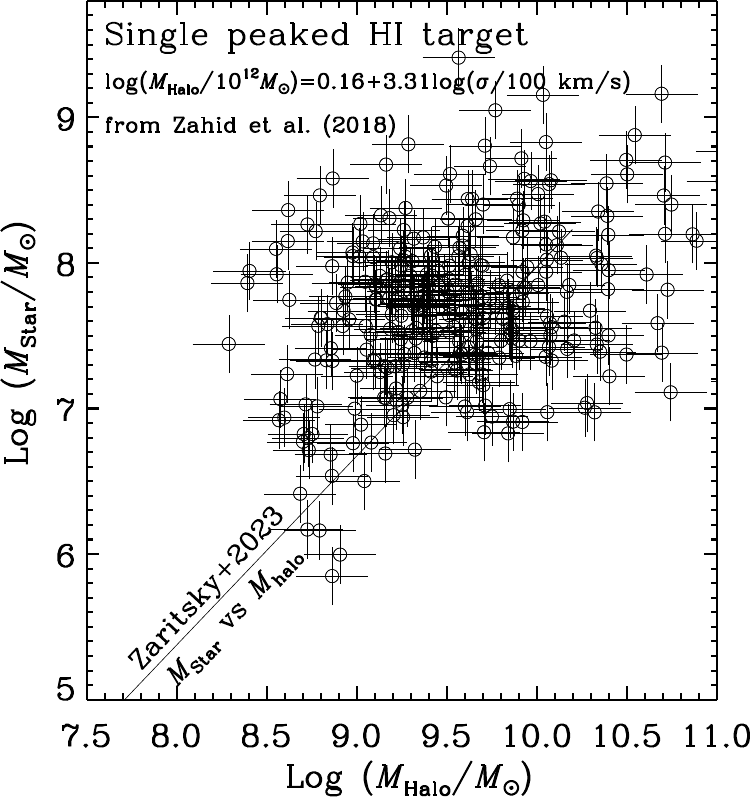}
    \caption{    {\bf Top panel:} Dynamical mass ($M_{\rm dyn}(<R_{\rm HI})$) vs stellar mass for the single- (red) and double- (blue) peaked \HI\ targets. The solid line is the $M_{\rm Star} vs M_{\rm halo}$ from \citet{2006ApJ...642L..37Z}. We do not transfer the dynamical mass within \HI\ into the total halo mass because of the potential uncertainty. The dynamical mass is consistently lower than the expected halo mass. {\bf Bottom panel:} Halo mass vs stellar mass for the single-peaked \HI\ targets. The halo masses are estimated from the formula given by \citet{2018ApJ...859...96Z}. 
    }
    \label{mhalo}
\end{figure}

\subsection{Outflow as an origin of the Velocity Dispersion}\label{sigma}

\HI-selected dwarf galaxies predominantly lie on or below the star-forming main sequence \citep[e.g.,][]{2012AJ....143..133H}, with typical star formation rates as low as $\sim10^{-2}\,M_\odot\,\mathrm{yr}^{-1}$. Previous studies comparing stellar and gas velocity dispersions suggest that for galaxies with $\mathrm{SFR} < 1\,M_\odot\,\mathrm{yr}^{-1}$, outflow velocities are generally not strongly dependent on the SFR \citep{2016A&A...588A..41C}. Empirical studies of outflow velocity as a function of SFR \citep[e.g.,][]{2015ApJ...811..149C, 2023ApJ...951..105D, 2024arXiv240608561T} find that galaxies with SFRs in the range of $10^{-2} - 10^{-1}\,M_\odot\,\mathrm{yr}^{-1}$ typically exhibit outflow velocities between 10 and 50 $\mathrm{km\,s^{-1}}$, which is comparable to the observed \HI\ profile widths in single-peaked systems (Figure~\ref{masssize}). Although the spatial extent of the \HI\ gas is much larger than that of the star-forming regions, low-level star formation may still contribute to the overall \HI\ line width, possibly contributing to the large scatter observed in the Faber–Jackson relation and the fundamental plane. A more detailed comparison between stellar absorption features (e.g., Na~I; \citealt{2010AJ....140..445C}) and the \HI\ velocity profiles could offer new insights into the role of neutral outflows in shaping the observed gas kinematics.

It is important to emphasize, however, that outflow properties are also strongly influenced by the gravitational potential of the dark matter halo. In the following section, we estimate the dark matter halo masses for our sample.

\subsection{From \HI\ Kinematics to Dark Matter Halo Dynamical Mass}

As discussed in Sections~\ref{sec32} and \ref{sec33}, we divide the \HI\ dwarf galaxies into two groups based on the shape of their \HI\ profiles. These two populations appear to follow the Tully–Fisher and Faber–Jackson relations, respectively, suggesting that their \HI\ kinematics might be able to be used to estimate the dynamical mass of their host dark matter halos.

Reliable estimates of the dark matter halo mass ($M_{\rm halo}$) are typically obtained from spatially resolved velocity maps modeled with dark matter halo profiles assumption \citep[e.g.,][]{2021ApJ...909...20S}. However, since our \HI\ observations do not provide spatially resolved kinematic maps, we estimate the dynamical mass ($M_{\rm dyn}$) using the \HI\ diameter inferred from an empirical scaling relation: $\log D_{\rm HI}\,[\mathrm{kpc}] = 0.506 \log(M_{\rm HI}/M_\odot) - 3.293$, where $D_{\rm HI} = 2R_{\rm HI}$ denotes the diameter of the region with \HI\ surface density $\Sigma_{\rm HI} > 1\,M_\odot\,\mathrm{pc}^{-2}$ \citep{2016MNRAS.460.2143W, 1997A&A...324..877B}. For galaxies with single-peaked profiles, we treat the \HI\ linewidth as a proxy for velocity dispersion and estimate the dynamical mass $M_{\rm dyn} = 4\,(W50/2.355)^2 \,R_{\rm HI}/G$ \citep{2010MNRAS.406.1220W}, while for double-peaked systems, we interpret it as tracing rotational velocity, and estimate the dynamical mass $M_{\rm dyn} = (W50^{\rm incl. corr.}/2)^2 R_{\rm HI}/G$.

Figure~\ref{mhalo} presents the inferred dynamical masses and halo masses. We find that galaxies with single-peaked \HI\ profiles tend to have similar stellar masses but lower halo masses compared to their double-peaked counterparts. This implies a lower $M_*/M_{\rm halo}$ ratio and highlights a large scatter in this quantity at the low-mass end. The inferred low halo masses of the single-peaked systems are comparable to those of UDGs reported to be deficient in dark matter \citep{2018Natur.555..629V, 2019ApJ...874L...5V}. It is therefore plausible that some of the single-peaked \HI\ galaxies in our sample are also deficient in dark matter.

As shown in Figure~\ref{massdensity}, both UDGs and our \HI-selected dwarfs exhibit similar stellar masses but UDGs have lower stellar mass surface densities. If the single-peaked systems indeed reside in shallower potential wells due to reduced halo masses, their stellar systems may be less gravitationally bound, potentially leading to more diffuse morphologies. Most \HI-detected UDGs have \HI\ masses above $10^8\,M_\odot$ \citep{2024A&A...681A..15M} and thus fall outside our sample selection. If lower-mass UDGs are intrinsically gas poor, the trends seen in stellar surface density (Figure~\ref{massdensity}) and halo mass (Figure~\ref{mhalo}) may trace an evolutionary pathway from \HI-rich dwarf galaxies to gas-poor dE galaxies, and eventually to UDGs.

\section{Summary}

We present a sample of HI-rich dwarf galaxies with $M_{\rm HI} < 10^8\,M_\odot$, selected from the first data release of the FASHI project. Optical counterparts are identified from DECaLS images and validated using scaling relations. We find that the stellar mass densities of these \HI\ dwarf galaxies are comparable to those of dwarf elliptical galaxies in low-redshift galaxy clusters. Compared to other dwarf galaxy populations, the \HI-selected dwarfs exhibit higher stellar mass densities than ultra-diffuse galaxies, and similar densities to \HI-selected low surface brightness galaxies, albeit with lower stellar masses. 

To investigate the dynamical state of these dwarf galaxies, we classify them into two groups based on their \HI\ profile shapes: single peaked and double peaked. We find that the double-peaked \HI\ targets follow the baryonic Tully–Fisher relation, as expected for rotation-dominated systems such as disk galaxies. In contrast, single-peaked systems are offset from the relation, which may be due to their dispersion-dominated kinematics or a deficiency in dark matter halo mass.

Further analysis reveals that single-peaked \HI\ dwarfs follow the trends of Faber–Jackson relation and fundamental plane of massive elliptical galaxies. This suggests that the relationship between stellar mass and halo mass in velocity-dispersion-dominated systems may be consistent across both low- and high-mass regimes. These results also imply that the \HI\ profile shape can be used as a proxy to estimate velocity dispersion in \HI\ single-peaked systems. Dynamical masses estimated from the \HI\ profiles indicate that single-peaked systems tend to reside in lower-mass halos, consistent with findings that some UDGs are deficient in dark matter. Future spatially resolved comparisons between stellar and \HI\ kinematics will be crucial for further constraining the dark matter halo properties of these systems.

While our sample selection focuses on the lowest \HI mass detections, it may miss some low-stellar-mass but \HI-rich, rotation-dominated galaxies. The FASHI survey is still ongoing, and a more comprehensive investigation of the \HI\ and optical properties of such systems will be carried out in the upcoming FASHI Survey DR2.

\begin{acknowledgments}

We sincerely thank the referee for the thoughtful and timely feedback, which has helped us significantly improve the clarity and robustness of the paper.
This work is sponsored (in part) by the Chinese Academy of Sciences (CAS) through a grant to the CAS South America Center for Astronomy. C.C. acknowledges NSFC grant No. 11803044 and 12173045. This work is supported by the China Manned Space Program with grant no. CMS-CSST-2025-A07. C.C. is supported by Chinese Academy of Sciences South America Center for Astronomy (CASSACA) Key Research Project E52H540301. We acknowledge support from the National Key Research and Development Program of China (grant No. 2023YFA1608100), and from the NSFC (grant Nos. 12122303, 11973039). 

This work made use of the data from FAST (Five-hundred-meter Aperture Spherical radio Telescope). FAST is a Chinese national mega-science facility, operated by National Astronomical Observatories, Chinese Academy of Sciences.

The DESI Legacy Imaging Surveys consist of three individual and complementary projects: the Dark Energy Camera Legacy Survey (DECaLS), the Beijing-Arizona Sky Survey (BASS), and the Mayall z-band Legacy Survey (MzLS). DECaLS, BASS and MzLS together include data obtained, respectively, at the Blanco telescope, Cerro Tololo Inter-American Observatory, NSF’s NOIRLab; the Bok telescope, Steward Observatory, University of Arizona; and the Mayall telescope, Kitt Peak National Observatory, NOIRLab. NOIRLab is operated by the Association of Universities for Research in Astronomy (AURA) under a cooperative agreement with the National Science Foundation. Pipeline processing and analyses of the data were supported by NOIRLab and the Lawrence Berkeley National Laboratory (LBNL). Legacy Surveys also uses data products from the Near-Earth Object Wide-field Infrared Survey Explorer (NEOWISE), a project of the Jet Propulsion Laboratory/California Institute of Technology, funded by the National Aeronautics and Space Administration. Legacy Surveys was supported by: the Director, Office of Science, Office of High Energy Physics of the U.S. Department of Energy; the National Energy Research Scientific Computing Center, a DOE Office of Science User Facility; the U.S. National Science Foundation, Division of Astronomical Sciences; the National Astronomical Observatories of China, the Chinese Academy of Sciences and the Chinese National Natural Science Foundation. LBNL is managed by the Regents of the University of California under contract to the U.S. Department of Energy. The complete acknowledgments can be found at https://www.legacysurvey.org/acknowledgment/.

The Siena Galaxy Atlas was made possible by funding support from the U.S. Department of Energy, Office of Science, Office of High Energy Physics under Award Number DE-SC0020086 and from the National Science Foundation under grant AST-1616414.

\end{acknowledgments}

\begin{contribution}

Cheng Cheng conceived the core idea of this study and carried out the primary measurements and detailed analysis. Ming Zhu and Chuan-Peng Zhang provided the FASHI dataset and played an essential role in data reduction and interpretation. All authors actively participated in project discussions, contributing to the refinement of the research questions, methodology, and scientific interpretation.



\end{contribution}

%
\facilities{FAST, KPNO:Mayall (Mosaic-3), Steward:Bok (90Prime), CTIO:Blanco (DECam)}

\software{astropy \citep{2013A&A...558A..33A,2018AJ....156..123A,2022ApJ...935..167A},  
          SWARP \citep{2002ASPC..281..228B},
          Source Extractor \citep{1996A&AS..117..393B},
          HiFAST \citep{2024SCPMA..6759514J}.
          }


\appendix

\section{Relations between the W50 or W20 with the rotation velocity for dwarf galaxies}\label{w50vrot}

To assess how good the approximation of using the line width (W50 or W20) as a proxy for the rotation velocity is, we collect several observations of dwarf galaxies with high-resolution \HI\, and show the results in Figure \ref{W50rot}. The high-resolution \HI\ data can reveal the rotation velocity, and the inclination is corrected by the \HI\ map. If we treat the $V_{\rm rot}$ from the \HI\ data cube as the real rotation velocity, in Figure \ref{W50rot} show that rotation velocity derived from W50 is roughly 5 km s$^{-1}$ larger, and W20 would have a much larger offset.

\setcounter{figure}{0}
\renewcommand{\thefigure}{A\arabic{figure}}

\begin{figure}[ht!]
    \centering
    \includegraphics[width=0.55\textwidth]{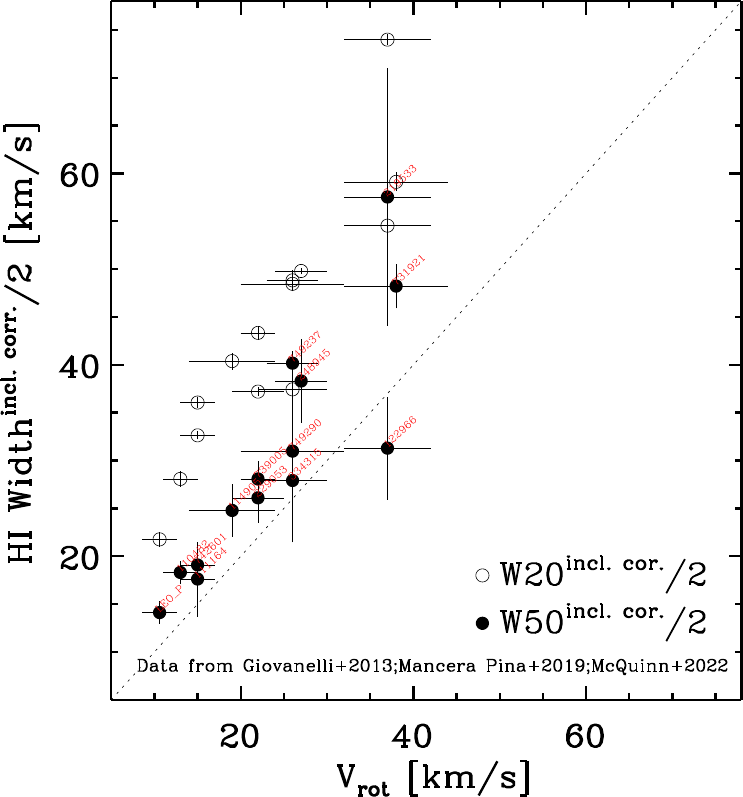}
    \caption{\HI\ line width from the global HI profile versus the rotation velocity ($V_{\rm rot}$). 
    The W50 (W20) are corrected by the inclination of the \HI\ image, and the $V_{\rm rot}$ is derived from the \HI\ data cube with HI spatial resolution. Therefore it is acceptable to make use of W50 to estimate the rotation velocity after inclination correction, since the offset is still less than the rotation velocity itself. In contrast, W20 exhibits an offset from $V_{\rm rot}$ that is comparable in magnitude to the rotation velocity itself, making it a less reliable proxy for the rotation velocity. IDs are adopted from the respective references \citep{2013AJ....146...15G, 2019ApJ...883L..33M, 2022ApJ...940....8M}.}
    \label{W50rot}
\end{figure}

\section{Stellar mass estimation of dwarf galaxies}\label{app:mstar_comparison}

Stellar mass estimation depends on the assumption of star formation history, metallicity, stellar population model, etc., and thus the mass-to-light-ratio method would include large scatter or bias. Since dwarf galaxies would have a less active star formation history and lower metallicity than massive galaxies, the mass-to-light ratio adopted in \citet{2003ApJS..149..289B} would bias the stellar mass estimation. To understand the scatter of our adoption, we compare our results with the mass-to-light ratio for irregular dwarf galaxies in \citet{2016AJ....152..177H}, and the results of low-surface-brightness galaxies in \citep{2020AJ....159..138D} in Figure \ref{mass-comp}. From Figure \ref{massdensity}, the \HI-dwarf galaxies are closer to the low-surface-brightness system, so we can expect that the results of \citet{2020AJ....159..138D} would have a more reliable stellar mass for the low surface brightness galaxies in our sample, while not all \HI-dwarf galaxies are low surface brightness galaxies. We also compare other results from \citet{2013MNRAS.430.2715I} and \citet{2009MNRAS.400.1181Z} in Figure \ref{mass-comp}. We can see a consistent trend between different methods with a bias about 0.2 dex at the low mass end. Meanwhile, SEDs from U to NIR bands of this dwarf galaxy sample are still needed to better constrain the stellar mass.

\setcounter{figure}{0}
\renewcommand{\thefigure}{B\arabic{figure}}

\begin{figure}[ht!]
    \centering
    \includegraphics[width=0.5\textwidth]{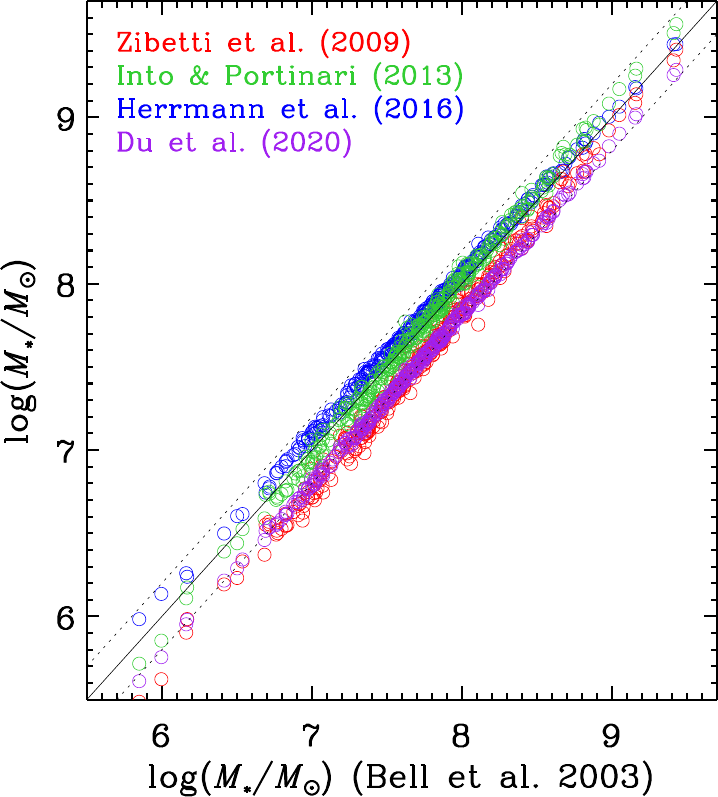}
    \caption{
    Stellar mass comparison between \citet{2003ApJS..149..289B} and the mass-to-light ratio from \citet{2016AJ....152..177H}, which is optimal to the irregular dwarf galaxies (blue circles) and the \HI-selected low-surface-brightness galaxies \citep{2020AJ....159..138D}. We also compare the results with other stellar mass estimators in \citet{2013MNRAS.430.2715I} and \citet{2009MNRAS.400.1181Z} with green and red circles. The solid and dotted lines are the 1:1 line and 0.2 dex.
    }
    \label{mass-comp}
\end{figure}

\setcounter{figure}{0}
\renewcommand{\thefigure}{C\arabic{figure}}

\begin{figure}
    \centering
\includegraphics[width=0.45\linewidth]{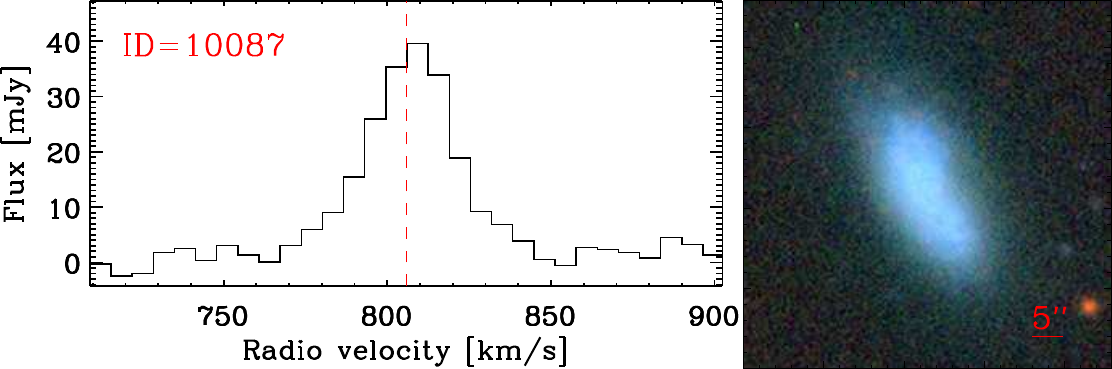}
\includegraphics[width=0.45\linewidth]{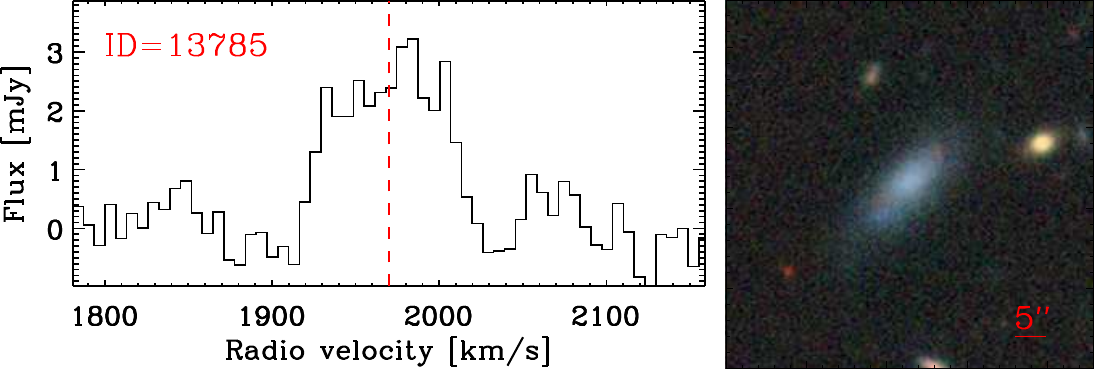}
\includegraphics[width=0.45\linewidth]{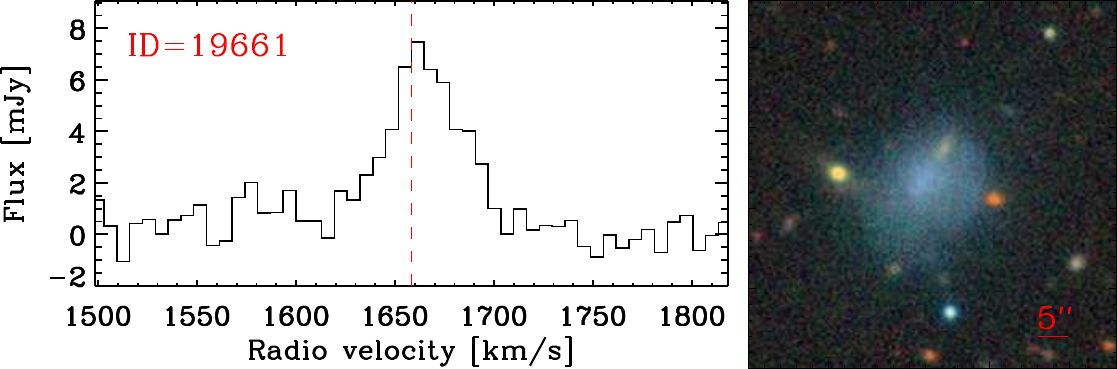}
\includegraphics[width=0.45\linewidth]{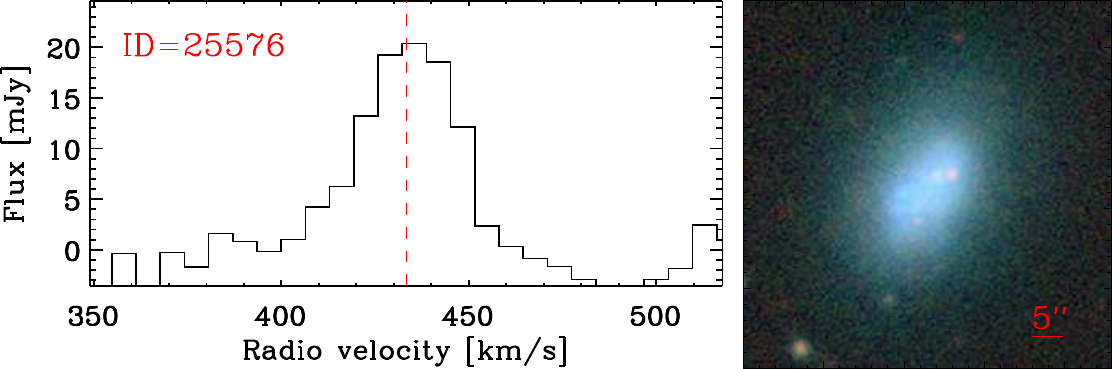}
\includegraphics[width=0.45\linewidth]{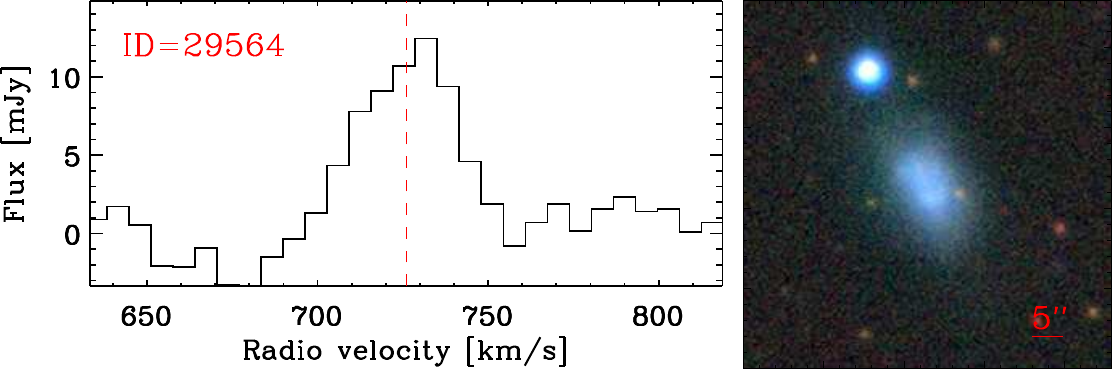}
\includegraphics[width=0.45\linewidth]{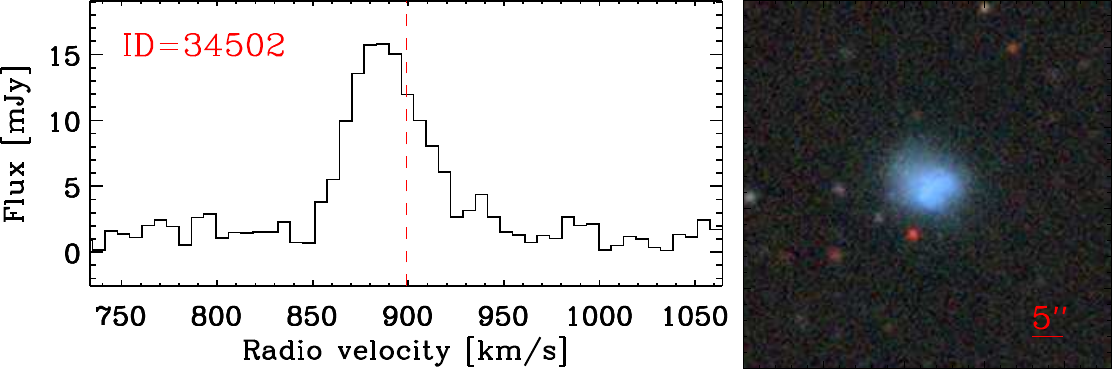}
\includegraphics[width=0.45\linewidth]{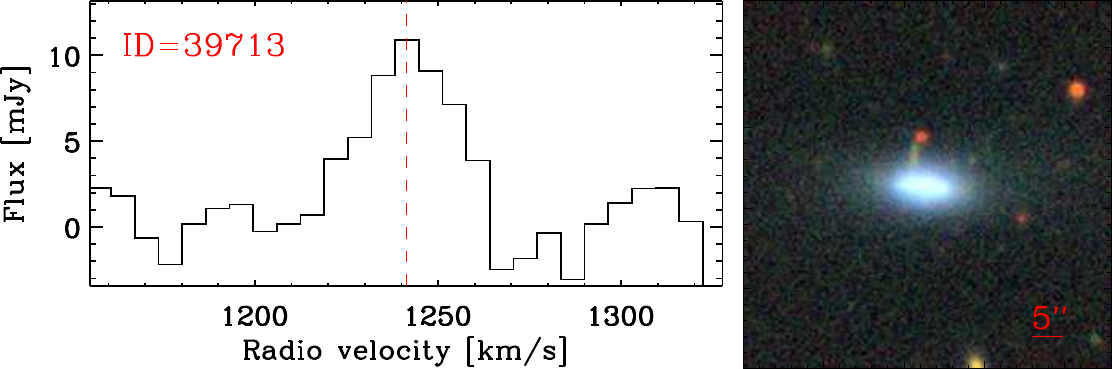}
\includegraphics[width=0.45\linewidth]{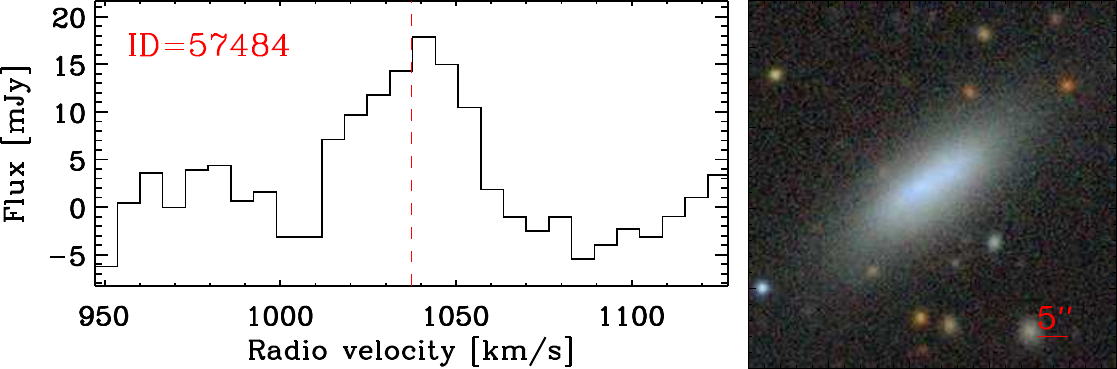}
\includegraphics[width=0.45\linewidth]{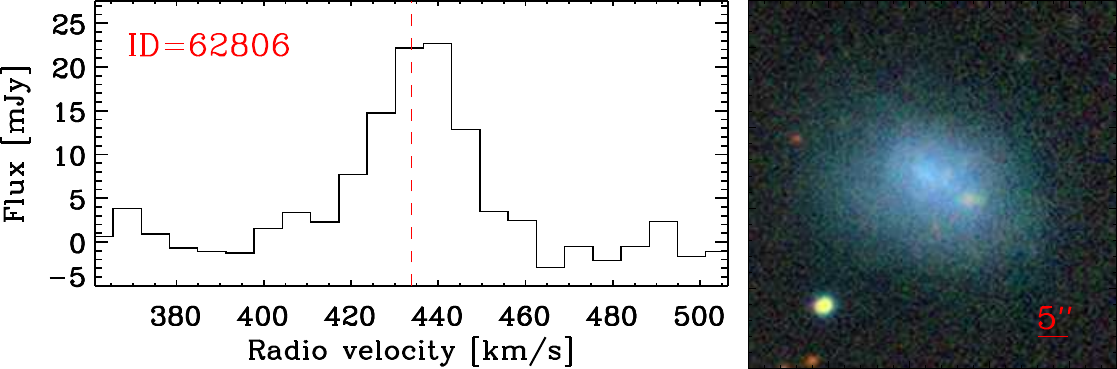}
\includegraphics[width=0.45\linewidth]{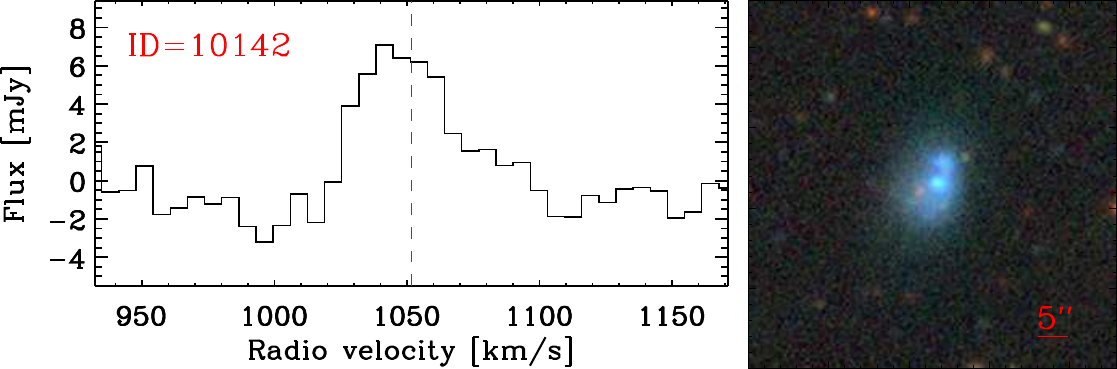}
\includegraphics[width=0.45\linewidth]{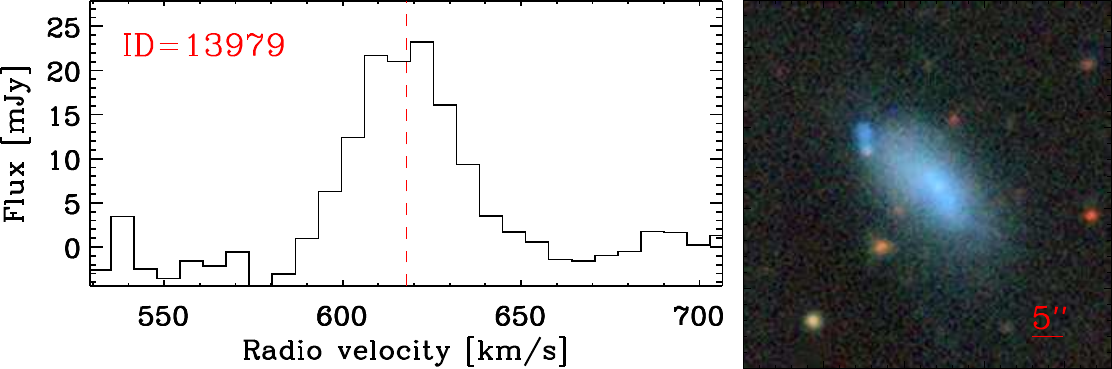}
\includegraphics[width=0.45\linewidth]{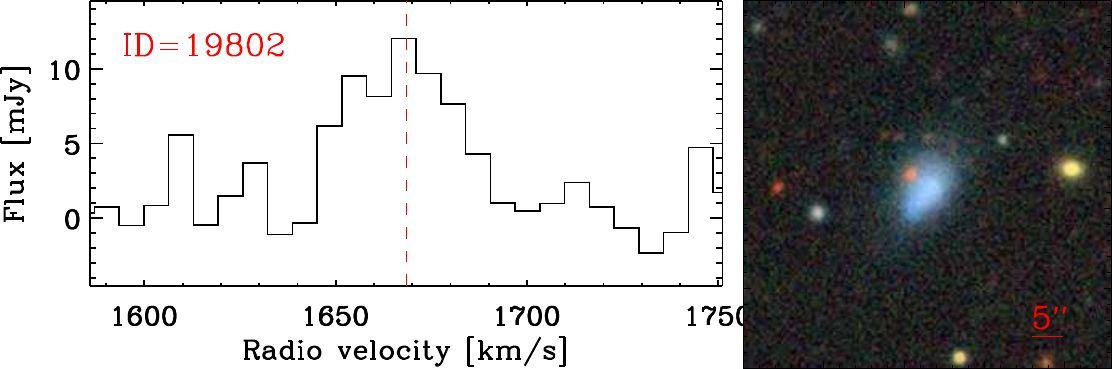}
\includegraphics[width=0.45\linewidth]{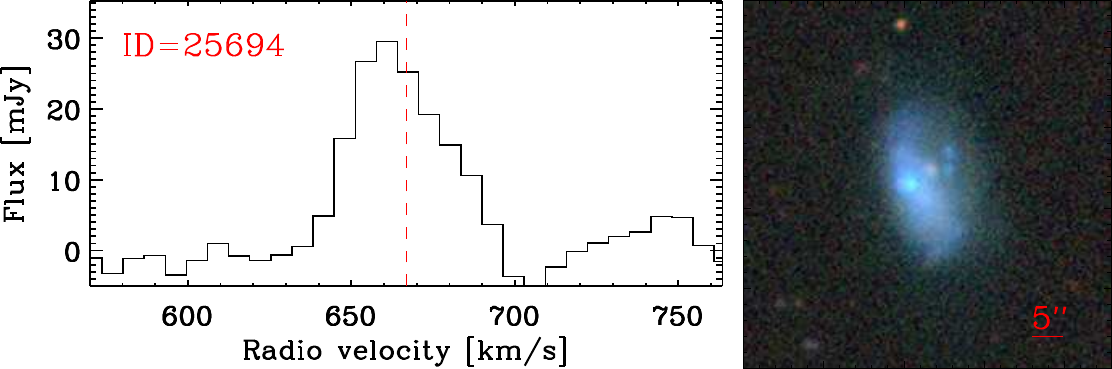}
\includegraphics[width=0.45\linewidth]{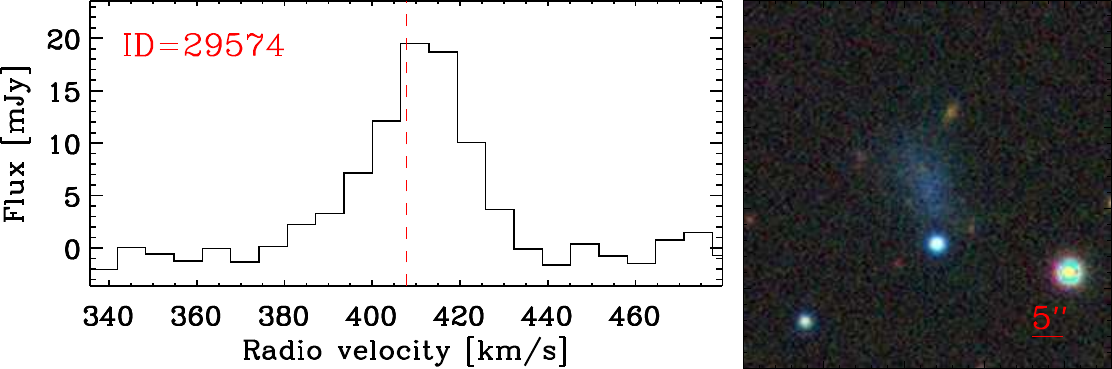}
\caption{Examples of the targets with single-peaked \HI\ profiles (left panels) and optical images of their counterparts ($g–r–i$ composites, $1'\times1'$) from the DECaLS survey (right panels). The channel width of the \HI\ spectrum is 6.4 $\rm km \, s^{-1}$. The complete figure set is available in the online journal.
}\label{SPHI}
\end{figure}

\begin{figure}
    \centering
\includegraphics[width=0.45\linewidth]{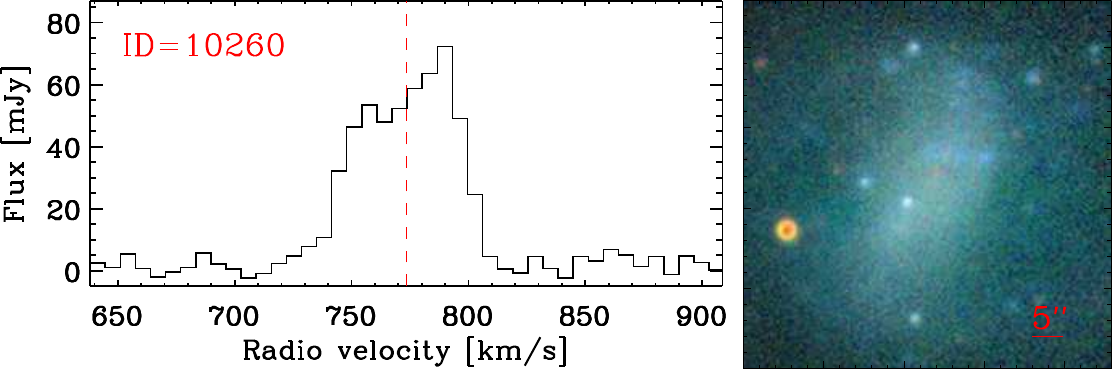}
\includegraphics[width=0.45\linewidth]{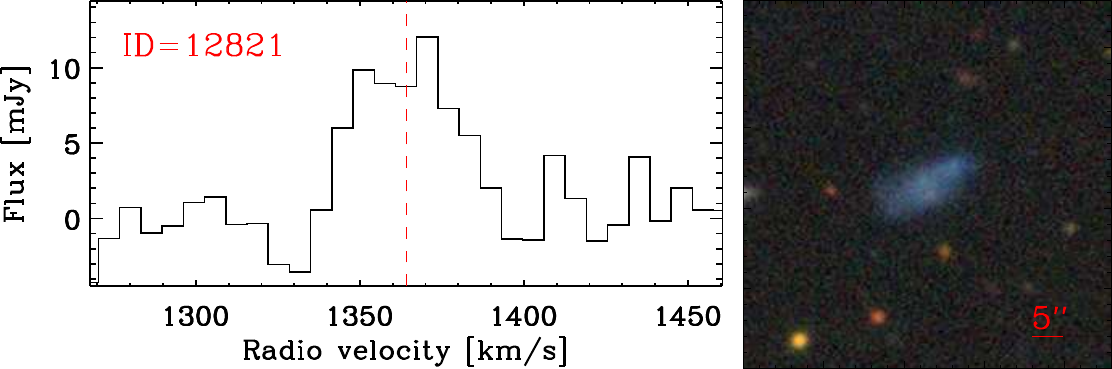}
\includegraphics[width=0.45\linewidth]{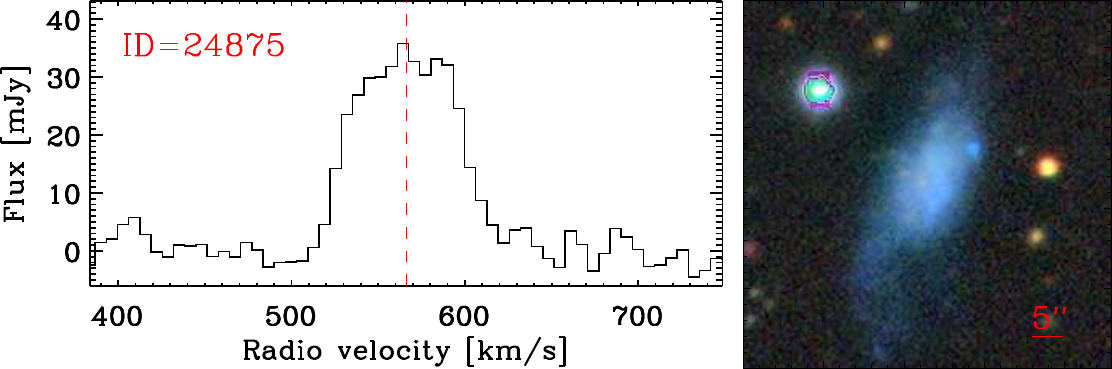}
\includegraphics[width=0.45\linewidth]{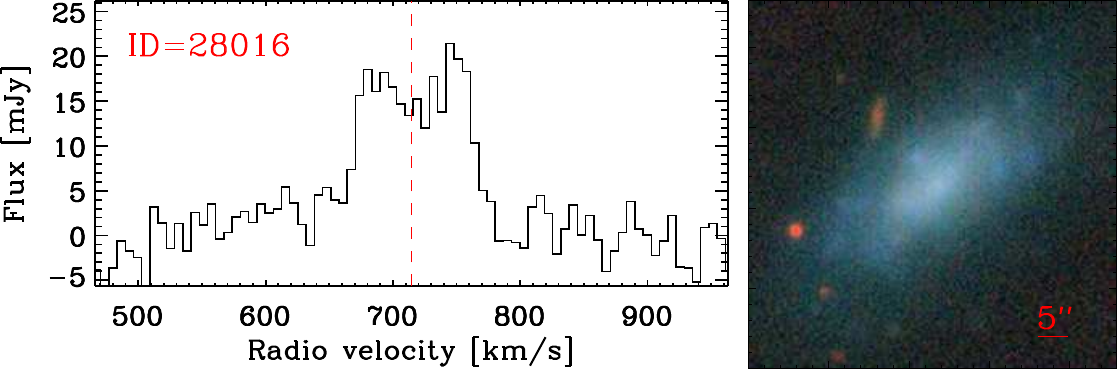}
\includegraphics[width=0.45\linewidth]{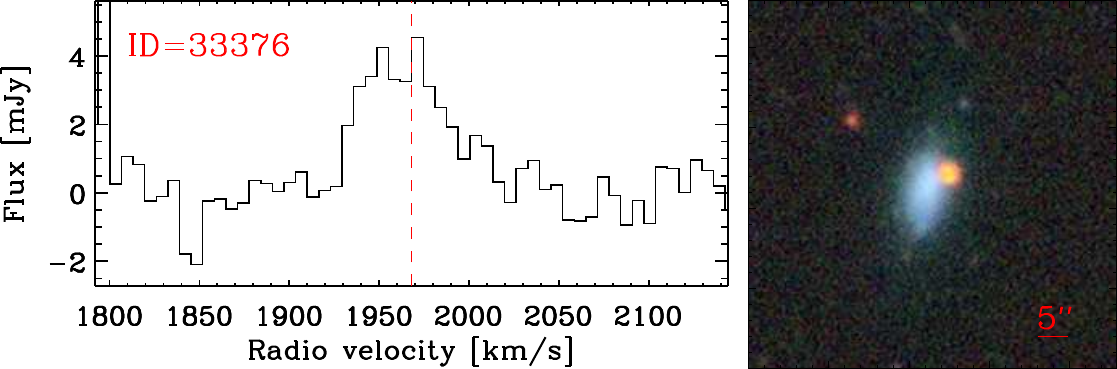}
\includegraphics[width=0.45\linewidth]{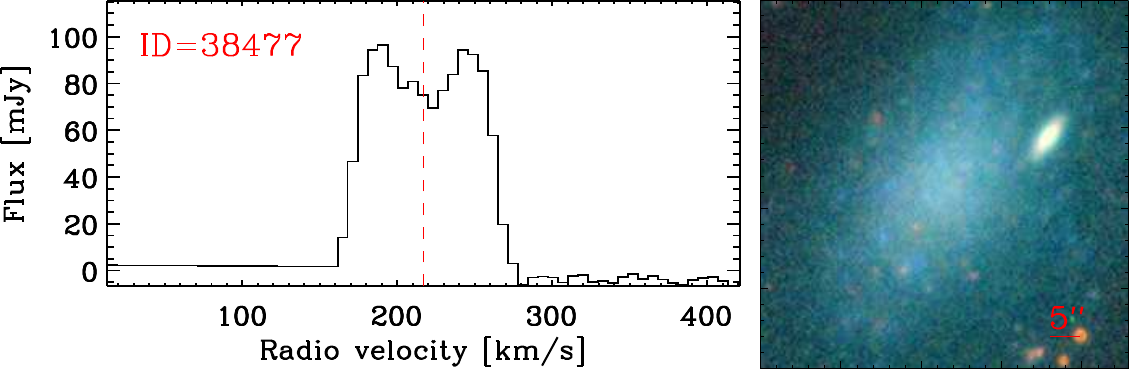}
\includegraphics[width=0.45\linewidth]{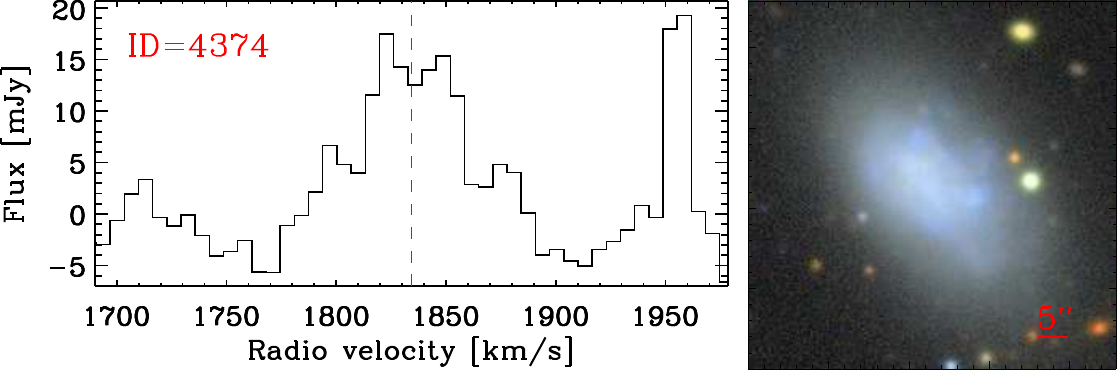}
\includegraphics[width=0.45\linewidth]{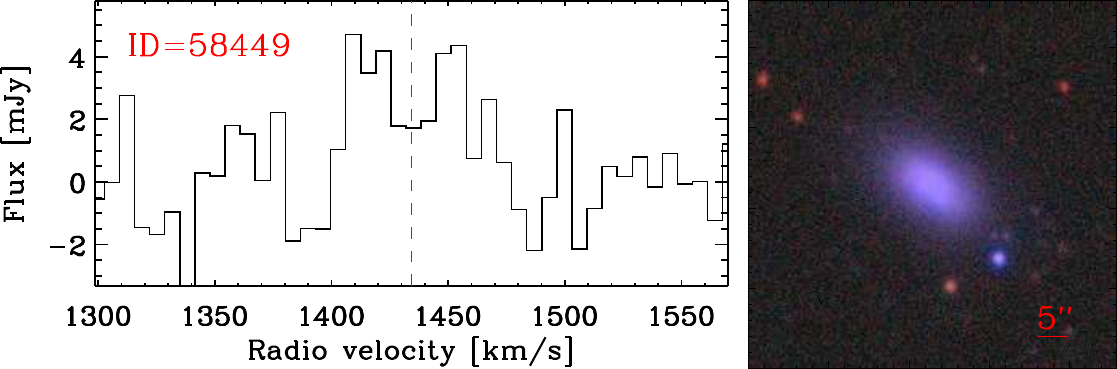}
\includegraphics[width=0.45\linewidth]{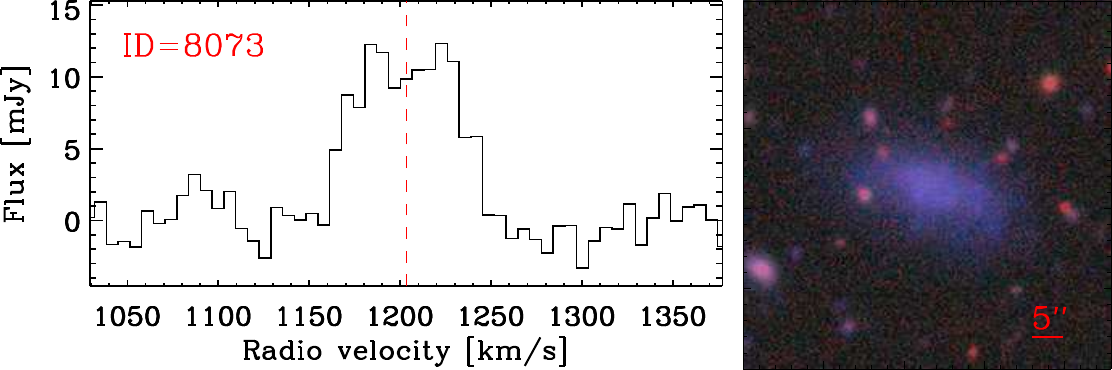}
\includegraphics[width=0.45\linewidth]{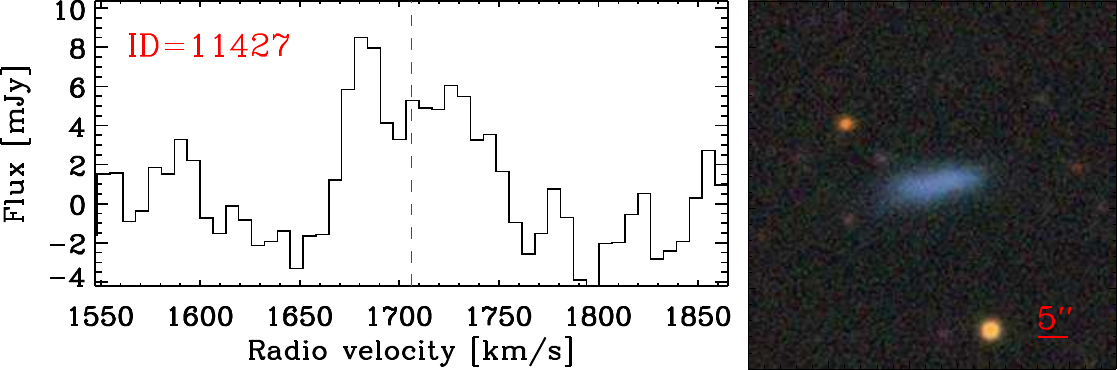}
\includegraphics[width=0.45\linewidth]{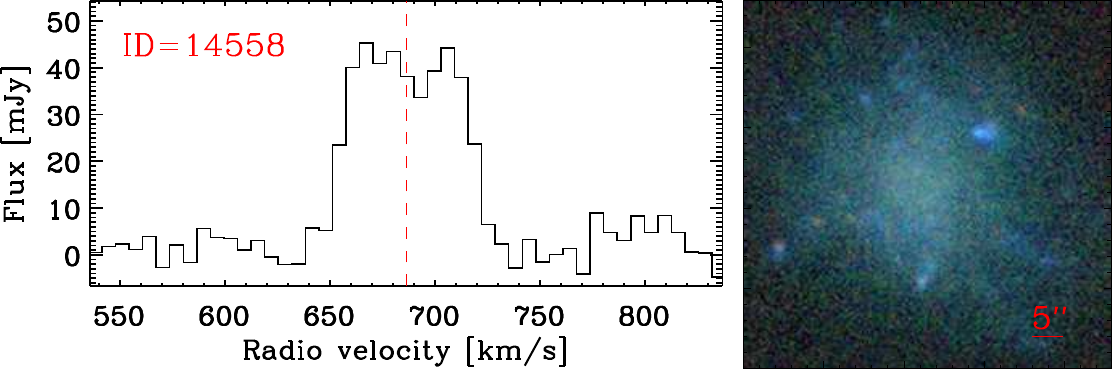}
\includegraphics[width=0.45\linewidth]{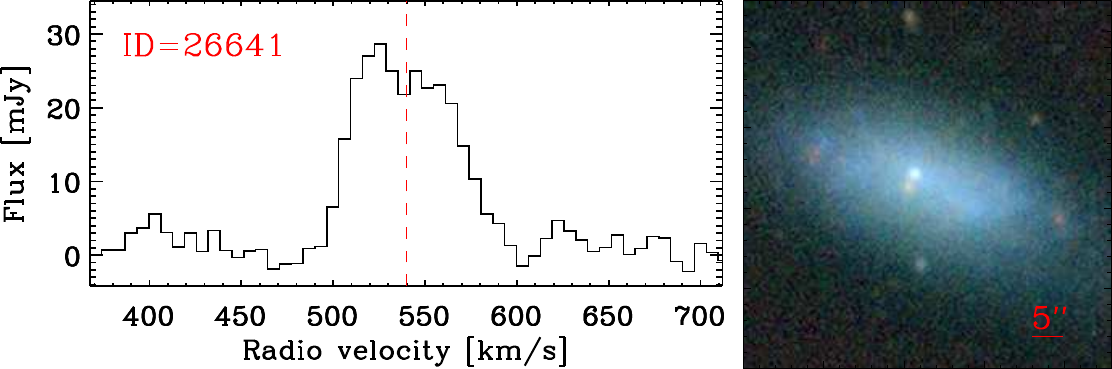}
\includegraphics[width=0.45\linewidth]{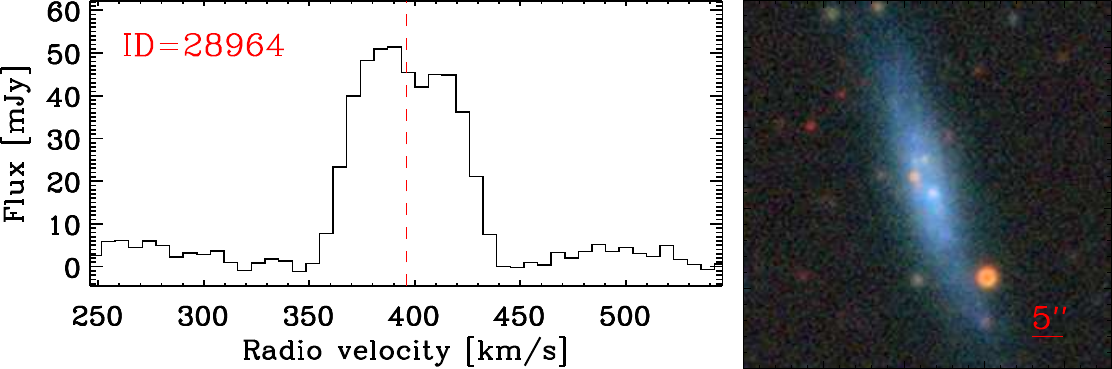}
\includegraphics[width=0.45\linewidth]{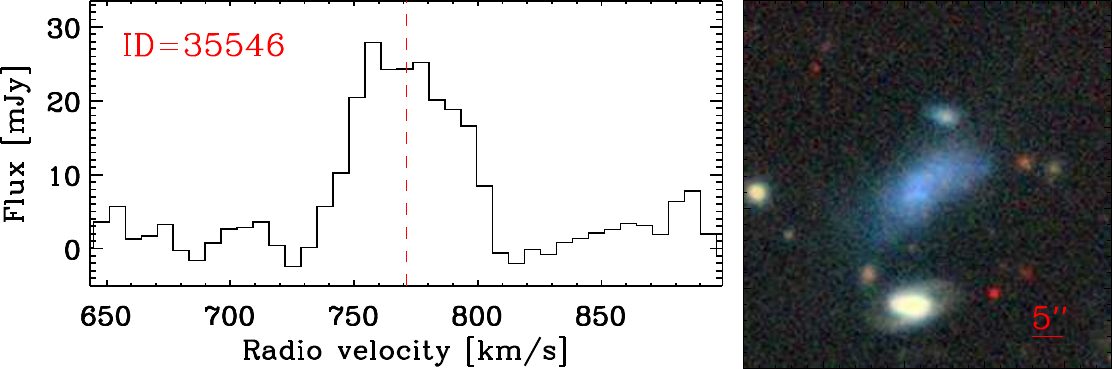}
\caption{Examples of the targets with double-peaked \HI\ profiles (left panels) and optical images of their counterparts ($g–r–i$ composites, $1'\times1'$) from the DECaLS survey (right panels). 
The channel width of the \HI\ spectrum 
is 6.4 $\rm km \, s^{-1}$. The complete figure set is available in the online journal.}
    \label{DPHI}
\end{figure}

\section{\HI\ spectra and optical images}

We present a randomly selected FAST \HI\ spectra alongside DECaLS three-color images (composed from $g$, $r$, and $z$ bands)in Figure \ref{SPHI}, which shows examples for targets with single-peaked \HI\ profiles, while Figure \ref{DPHI} illustrates those with double-peaked profiles. All spectra and images are available in the online figure sets.

\newpage
\startlongtable
\begin{deluxetable*}{lccccccc}
\digitalasset
\tablewidth{0pt}
\tablecaption{Catalog of optical counterparts for the low \HI\ mass sample.\label{tab:description}}
\tablehead{
\colhead{FASHI ID} & \colhead{OC R.A. (J2000)} & \colhead{OC Decl. (J2000)} & \colhead{c$z$ (km s$^{-1}$)} & \colhead{$m_r$ AB mag} & \colhead{$m_g$ AB mag} & \colhead{Re (arcsec)} & \colhead{HI profile flag}
}
\startdata
60012 & 13:49:27.6 & -06:05:18.1 & 1517 &    16.9599 $\pm$     0.0031 &    17.2280 $\pm$     0.0023 &        5.6  &  S \\
60054 & 12:53:31.6 & -05:55:40.4 & 1068 &    15.2933 $\pm$     0.0022 &    15.7946 $\pm$     0.0015 &       15.2  &  D \\
57486 & 13:01:10.9 & -05:33:24.6 & 1200 &    13.5354 $\pm$     0.0004 &    14.1124 $\pm$     0.0004 &        9.1  &  S \\
57485 & 13:01:05.2 & -05:28:19.2 & 1083 &    15.6423 $\pm$     0.0017 &    16.0505 $\pm$     0.0016 &        9.0  &  S \\
57499 & 09:56:44.3 & -05:08:26.4 & 1481 &    18.9305 $\pm$     0.0158 &    19.0504 $\pm$     0.0098 &        3.8  &  S \\
57498 & 11:31:45.0 & -05:07:33.8 & 1050 &    17.9549 $\pm$     0.0046 &    18.3567 $\pm$     0.0035 &        2.9  &  S \\
57484 & 12:47:13.5 & -05:07:02.3 & 1041 &    16.0296 $\pm$     0.0024 &    16.5795 $\pm$     0.0019 &        7.6  &  S \\
1206 & 11:27:43.6 & -04:55:30.5 & 968 &    15.2122 $\pm$     0.0015 &    15.4316 $\pm$     0.0007 &       14.4  &  D \\
\enddata
\tablecomments{HI profile flag: S for single-peaked HI profile. D for double-peaked HI profile. Only a portion of this table is shown here to demonstrate its form and content. A machine-readable version of the full table is available.}
\label{table1}
\end{deluxetable*}





\end{document}